\documentclass{aa}  

\usepackage{graphicx}
\usepackage{subcaption}
\usepackage{txfonts}
\usepackage{pdfpages}

\usepackage{natbib}
\bibpunct{(}{)}{;}{a}{}{,}
\usepackage{hyperref}

\newcommand{\gaia}{\textit{Gaia}\xspace}
\newcommand{\gdr}[1]{\gaia~DR#1\xspace}
\newcommand{\gedr}{\gaia~EDR3\xspace}
\newcommand{\bprp}{\ensuremath{G_{\rm BP}-G_{\rm RP}}\xspace}

\begin{document} 

   \title{Exploring the structure and kinematics of the Milky Way through A stars\thanks{The catalogue of A-type stars described in this paper is available via ...}}

   \author{J. Ard\`evol \inst{1,2,3}   \fnmsep
   \thanks{\email{jardevol@icc.ub.edu}}
          \and
          M. Mongui\'o \inst{1,2,3}
          \and
          F. Figueras \inst{1,2,3}
          \and
          M. Romero-G\'omez \inst{1,2,3}
          \and 
          J.M. Carrasco  \inst{1,2,3}
          }
   \institute{Institut de Ci\`encies del Cosmos (ICCUB), Universitat de Barcelona (UB), Mart\'i i Franqu\`es 1, E-08028 Barcelona, Spain
   \and
   Departament de Física Qu\`antica i Astrof\'isica (FQA), Universitat de Barcelona (UB), Mart\'i i Franqu\`es 1, E-08028 Barcelona, Spain
   \and
   Institut d'Estudis Espacials de Catalunya (IEEC), c. Gran Capit\`a, 2-4, 08034 Barcelona, Spain}

   \date{Received XXX; Accepted YYY}

  \abstract
   { Despite their relatively high intrinsic brightness and the fact that they are more numerous than younger OB stars and kinematically colder than older red giants, A-type stars have rarely been  used as Galactic tracers. They may, in fact, be used to fill the age gap between these two tracers, thereby allowing us to evaluate the evolutionary and dynamic processes underlying the transition between them.}
   {We analyse Galactic disc structure and kinematic perturbations up to 6~kpc from the Sun based on observations of A-type stars.}
   {This work presents a catalogue of A-type stars selected using the IGAPS photometric survey. It covers the Galactic disc within $30\,^{\circ} \leq l \leq 215\,^{\circ}$ and $|b| \leq 5\,^{\circ}$ up to a magnitude of $r \leq 19$~mag with about 3.5 million sources. We used {\it Gaia} Data Release 3 parallaxes and proper motions, as well as the line-of-sight velocities, to analyse the large-scale features of the Galactic disc. We carried out a study of the completeness of the detected density distributions, along with a comparison between the $b < 0\,^{\circ}$ and $b > 0\,^{\circ}$ regions. Possible biases caused by interstellar extinction or by the usage of some kinematic approximations were examined as well.}
   {We find stellar overdensities associated with the Local and the Perseus spiral arms, as well as with the Cygnus region. We find that A-type stars also provide kinematic indications of the Galactic warp towards the anticentre, which displays a median vertical motion of $\sim$~$6$-$7 \,\rm{km\,s^{-1}}$ at a Galactocentric radius of $R=14$~kpc. It starts at $R\!\approx\!12$~kpc, which supports the scenario where the warp begins at larger radii for younger tracers when compared with other samples in the literature. We also detect a region with downward mean motion extending beyond 2~kpc from the Sun towards $60\,^{\circ}\lesssim l \lesssim 75\,^{\circ}$ that may be associated with a compression breathing mode. Furthermore, A-type stars reveal very clumpy inhomogeneities and asymmetries in the $V_Z$-$V_{\phi}$ velocity space plane.}
   {}

   
   \titlerunning{Milky Way through A stars}
   \authorrunning{J. Ard\`evol et al.}
   
   \maketitle

\section{Introduction}
    Our knowledge of the Milky Way has been vastly improved in recent decades, but there are still many unknowns that remain to be deciphered. Our location within the Galaxy makes it more difficult to detect its structure with respect to external galaxies, although our advantageous position has allowed us to analyse our own Galaxy in much greater detail. In turn, this has enabled  studies of the density and kinematic peculiar patterns produced by non-axisymmetric features (e.g. spiral arms, the warp, bar resonances, etc.) or by external interactions (as the Magellanic Clouds or the Sagittarius dwarf galaxy).

    The {\it Gaia} mission \citep{GaiaMission_2016} is helping us learn more about the previously observed properties of the Milky Way, as well as to discover other non-equilibrium structures \citep[e.g.][and many others]{PhaseSpiral_Antoja2018, GaiaEnceladus_Belokurov2018, GaiaEnceladus_Helmi2018}. As an example, \cite{AC_Teresa2021} and \cite{Disturbances_McMillan2022} have demonstrated that {\it Gaia} measurements are able to highlight large non-equilibrium features of the Milky Way. The huge amount of the high-quality data it has already produced makes it ideal for analysing the structure of our Galaxy. Nevertheless, on-ground photometric and spectroscopic surveys are also extremely valuable as a complement to the {\it Gaia} data. Some examples of these additional observations that are relevant to the present study are the photometry of IGAPS \citep{IGAPS_Monguio2020}, LAMOST spectra \citep{LAMOST_Cui2012}, and spectra  obtained by WEAVE \citep{WEAVE_Dalton2016, WEAVE_Jin2022} in the future.

    A multi-tracer study of the Milky Way allows us to disentangle its structure and kinematics within a wide range of ages and then to examine their evolution. Some of the more widely used Galactic tracers are interstellar gas and dust, open clusters, young OB stars, and old red giants, among others. Each of them displays its own individual properties, requires different observational methods, exhibits its own biases and, in general, presents a different view of the Milky Way.
    As a non-exhaustive list of tracers used to analyse the Galactic structure, neutral hydrogen (HI) detected with radio wavelengths (that are barely affected by dust extinction) traces several gaseous spiral arms up to more than 15~kpc from the Sun \citep[e.g.][]{1stHImap_Oort1958, HIspiralArms_Levine2006, AllGasArms_Nakanishi2016, HIspiralArms_Koo2017}. \cite{MasersPerseus_Sakai2019} and \cite{SpiralArms_Reid2019} used masers associated with zero-age O-type stars to study the distribution and the kinematics of the youngest components of the spiral structure. \cite{Structure_Poggio2021} used young tracers (upper main sequence stars, cepheids, and open clusters) to find high-density structures associated with spiral arms. \cite{CepSpur_Pantaleoni2021} used massive OB stars and stressed the presence of a coherent feature that they refer as Cepheus Spur. In comparison, \cite{OCvoid_CantatGaudin2020} constructed their own catalogue of open clusters, concluding that their distribution is highly dependent on their age, with the oldest ones presenting signatures of both Galactic warp and flare.
    The Galactic warp has been examined using several tracers such as HI gas \citep{HIwarp_Levine2006, HIspiralArms_Koo2017}, infrared emission from dust and cold giants \citep{IRwarp_Freudenreich1994}, and cepheids \citep{WarpCepheids_Chen2019, WarpCepheids_Skowron2020} as well as young OB and old Red Giant Branch (RGB) stars \citep{Warp_Merce2019}.

    Using tracers with different stellar ages is key to understand the evolution of the Milky Way \citep{Evolution_Amores2017}. Since intermediately young A-type stars are relatively intrinsically bright and numerous, they are suitable for tracing density structures in the Galactic disc. They have typical masses of about $3M_{\odot}$ \citep{Aproperties_Griv2020} and ages of the order of 0.3-1.0~Gyr \citep{SpiralPotentialMW_Grosbol2018}. They are young enough to have small velocity dispersion \citep[around $20\,\rm{km\,s^{-1}}$ according to][]{VelocityDispersionFig2_Aumer2009, AFpencilbeams1_Harris2018, AFpencilbeams2_Harris2019}, while at the same time being old enough to have orbited the Galaxy up to four times and so, to have interacted gravitationally with the potential of different Galactic structures. Therefore, they are also appropriate for kinematic studies. In fact, these tracers open a new window to understand the Milky Way disc by offering the best of the two most frequently used types of tracers; namely, the brightness of giants and upper main sequence stars, the low velocity dispersion of young stars and the relatively high abundance of cooler stars. A few examples of research already performed with these stars are the following ones. \cite{selection_Drew2008} analysed the spatial distribution of A-type stars in the OB association Cyg OB2, while \cite{StructureAstarsIPHAS_Sale2010} studied the Galactic disc stellar density profile in a region of $40\,^{\circ}$ in Galactic longitude around the anticentre (AC). Both papers showed the potential of selecting A-type stars using the $(r-H\alpha)$ vs $(r-i)$ colour-colour diagram. \cite{PerArmDetection_Monguio2015} detected the stellar overdensity associated with the Perseus arm for the first time using B4-A1 stars and located it at $1.6\pm0.2$~kpc from the Sun. \cite{SpiralPotentialMW_Grosbol2018} compared B- and A-type stars kinematics near the Galactic centre (GC) with simulations and concluded that the Milky Way potential has two major arms. \cite{AFpencilbeams1_Harris2018, AFpencilbeams2_Harris2019} analysed the rotation curve of the Milky Way using the 6D configuration space (positions and velocities) of A- and F-type stars, but only in two specific lines of sight.
    
    This paper is structured as follows. Section~\ref{sectionDefinitions} defines reference systems used throughout the work, while Sect.~\ref{sectionData} describes the data and different samples used. The stellar distribution in the XY plane and in the vertical direction are studied in Sect.~\ref{sectionStructureResults}. On the other hand, Sect.~\ref{sectionKinematics} describes a wide kinematic analysis of the samples. Section~\ref{sectionDiscussion} presents a discussion of our results and comparison  with the literature. Finally, Sect.~\ref{sectionConclusions} summarises the findings and conclusions of the work.

\section{Coordinate systems}\label{sectionDefinitions}
    Several reference systems are used across this paper. In this section, we define them, along with the notation and the reference values used.
    
    The spherical Galactic system is centred at the Sun. It is described by $(l,\,b,\,d)$, where the first two variables correspond with Galactic longitude and Galactic latitude, respectively, and $d$ stands for the distance between a given star and the Sun. Proper motions in these directions are referred to as $(\mu_{l},\,\mu_{b})$, while $v_{\rm los}$ are the line-of-sight velocities. Projection effects are already considered in the sense that $\mu_{l}$ contains the ${\rm cos}(b)$ term. Velocities in these Galactic directions are called $(v_l^{\rm corr}, \,v_b^{\rm corr})$ and are computed as:
    \begin{equation}\label{v_uncorr}
        v_{\gamma}^{\mathrm{uncorr}} = 4.7404705 \, d \, \mu_{\gamma} \mathrm{\ \ with} \ \gamma=l,b,
    \end{equation}
    where $v_{\gamma}^{\rm uncorr}$ is in $\rm{km\,s^{-1}}$ provided that $d$ is in kpc and $\rm{\mu_{\gamma}}$ is in $\rm{mas\,yr^{-1}}$. Both velocities given by Eq.~(\ref{v_uncorr}) can be corrected for solar motion according to:
    \begin{eqnarray}
    \nonumber
        \label{eq:v_corr}
        v_l^{\rm{corr}} &=& v_l^{\rm{uncorr}} - U_{\odot}{\rm sin}(l) + V_{\odot}{\rm cos}(l), \\
        v_b^{\rm{corr}} &=& v_b^{\rm{uncorr}} - U_{\odot}{\rm cos}(l){\rm sin}(b) \\ \nonumber &&-V_{\odot}{\rm sin}(l){\rm sin}(b) + W_{\odot}{\rm cos}(b),
    \end{eqnarray}
    where $(U_{\odot}, \, V_{\odot},  \, W_{\odot})$ are the components of the solar motion: $V_{\odot}$ includes both the circular velocity of the local standard of rest and the peculiar azimuthal velocity of the Sun with respect to it \citep{Thesis_Hoda2015, AFpencilbeams2_Harris2019, AC_Teresa2021}. We use the solar motion from \cite{Usun_DrimmelPoggio2018} [$U_{\odot}$] and \cite{VWsun_Reid2020} [$V_{\odot}$ and $W_{\odot}$], being $(U_{\odot}, \, V_{\odot}, \, W_{\odot})=(9.5, \, 250.7, \, 8.56)\,\rm{km\,s^{-1}}$ once scaled to the solar radius ($R_{\odot}$).

    Heliocentric Cartesian coordinates are named as $X$, $Y$, and $Z$, whereas Galactocentric ones are distinguished with a subscript: $X_{\rm Gal}$, $Y_{\rm Gal}$, and $Z_{\rm Gal}$. They transform as $X=X_{\rm Gal}+R_{\odot}$ and $Z=Z_{\rm Gal}+Z_{\odot}$, while the Y component of both reference systems coincide.
    The used distance between the Sun and the GC (i.e. the solar radius) is equal to $R_{\odot}=8.249 \pm 0.009$~kpc \citep{Rsun_Gravity2020} while the solar vertical coordinate $Z_{\odot}$ is assumed to be zero as a simplifying approximation (thus, $Z=Z_{\rm Gal}$). X increases towards the GC in the heliocentric case or away from the Sun in the Galactocentric one. In either case, Y grows in the direction that rotation has at $X_{\rm Gal}<0$, and Z towards the north Galactic pole (NGP).
    
    The cylindrical Galactocentric radial, azimuthal, and vertical coordinates are referred to as $R$, $\phi$, and $Z_{\rm Gal}$; while their respective velocities are $V_R$, $V_{\phi}$, and $V_Z$. Also, $(R,\,\phi,\,Z_{\rm Gal})$ define a left-handed system, in which $R$ increases away from the GC and $\phi$ originates at the Sun-AC direction, increasing clockwise as seen from the NGP.
    In the case of the three velocities: $V_R$ is positive outwards; $V_{\phi}$, in the direction of rotation (clockwise from the NGP); and $V_Z$, towards the NGP.

\section{Data}\label{sectionData}
    The sample of A-type stars was selected using IGAPS photometric bands $i$, H$\alpha$, and $r$ from IPHAS \citep{IPHAS_Drew2005}, which are centred at 774.3, 656.8, and 624.0~nm, respectively. It covers the northern Galactic plane within $l \in [30 \, , \, 215] \,^{\circ}$ and $|b| \leq 5 \,^{\circ}$. We used the $(r-H\alpha)$ vs $(r-i)$ colour-colour diagram and a similar procedure to that described in \cite{selection_Drew2008}. The first step to obtain the working sample was to remove noise-like sources forcing the variable Class defined in \cite{IGAPS_Monguio2020} to be different from 0. White dwarfs (WD) and supergiants were avoided by imposing $-0.1$~mag$<(r-i)<2.2$~mag. Then, following \cite{IGAPS_Monguio2020}, A-type stars were selected using the A0-A5 sequence line according to:
    \begin{eqnarray}
        \label{eq:select}
        (r-H\alpha)-(\delta r-\delta H\alpha) -[0.0032 +0.3735(r-i) \\
        -0.0608(r-i)^2  +0.0041(r-i)^3]&<&0, \nonumber
    \end{eqnarray}
    with the $(\delta r-\delta H\alpha)$ term accounting for the uncertainties in the observed $(r-H\alpha)$ colour. Photometric errors increase sample contamination specially for faint stars. For this reason, we limited the previous selection at $r \leq 19$~mag, which led to a sample with 3\,532\,751 stars. This sample is available through CDS as a table having the columns described in Appendix~\ref{appendixColumns}.
    
    Using a testing sample in the AC with near 2.05·$10^{5}$ stars of any spectral type that have information both in IGAPS and in LAMOST DR8\footnote{\url{http://www.lamost.org/dr8/v2.0/catalogue}.} catalogues, we found that 23\% of stars selected as A0-A5 by the IGAPS photometric selection defined above are classified as other spectral types according to LAMOST and only 10\% are not classified as types A0-A9. However, around half of that 23\% sample have at least one LAMOST signal-to-noise ratio lower than 2. All this confirms that the contamination in our selected sample is small, well below 20\%. This estimation agrees with \cite{StructureAstarsIPHAS_Sale2010} and \cite{AFpencilbeams1_Harris2018}, who found about 10-20\% of contamination in their selections, which were made using similar methodologies.

    We also compared our selection with the golden sample of OBA stars from \cite{goldOBA_Creevey2022}, restricted to a common sky region. Despite not including OB stars, our sample contains many more targets, and reaches both deeper distances and limiting magnitudes (with 90th percentiles of {\it Gaia} $G$ passband being 16.5~mag for the golden sample and 18.8~mag for our A-type stars). On the other hand, the {\it Gaia} golden sample includes few stars closer and apparently brighter that are not included in our sample due to saturation in the IGAPS photometric catalogue.

    Once the sample was selected, parallaxes ($\varpi$) and proper motions were then obtained from {\it Gaia} Data Release 3 \citep[DR3,][]{GaiaDR3_2022}. To do this, the sample was cross-matched with this catalogue utilising a 1" radius. This procedure gave 3\,512\,224 counterparts (99.4\% of our initial sample), out of which 3\,490\,765 (98.8\%) have the three aforementioned quantities. The number of sources having ${\rm RUWE}<1.4$ \citep{ruwe_Lindegren2020} is 3\,300\,599 (94.5\% of stars with astrometry). We are also interested in the line-of-sight velocities. There are 31\,934 sources (0.9\% of those with astrometry) with {\it Gaia} DR3 $v_{\rm los}$ available. It is also worth noting that they have a limiting magnitude of $G_{\rm RVS}=14$~mag. These line-of-sight velocities were corrected according to \cite{radVelCorrectionDR3_Katz2022}, which provided a small correction always below $0.4\,\rm{km\,s^{-1}}$ for our sample. The correction described by \cite{radVelCorrectionDR3notapplied_Blome2022} was not applied since most of the stars did not match the requirements for it.

    Heliocentric distances were computed with the exponentially decreasing space density (EDSD) prior \citep{EDSD_BailerJones2015}, which rely on a very small number of assumptions and just a single parameter, namely, the length scale. After checking different values for this length scale, the value of 3~kpc seems to be more appropriate for a sample of A-type stars \citep[in agreement with the length scale of $3020\pm300$~pc derived by][]{StructureAstarsIPHAS_Sale2010}. We studied as well other distance estimators such as those provided by \cite{Dist_BailerJones2021}. First, the geometric distances whose prior relies on several parameters varying for different directions in the sky and second, their photogeometric distances that also incorporates photometry in its prior. The priors used to compute these distances are usually based on colder -- more numerous -- stars and are not so well-behaved for our relatively blue stars lying near the Galactic plane (where extinction effects can be relevant).
    
    Nevertheless, it should be taken into account that any distance estimator includes biases for parallaxes with large errors. In particular, the selected one tends to accumulate them at a distance equal to twice the length scale (i.e. at 6~kpc from the Sun in our case). In consequence, the presence of any buildup at this particular distance should be suspicious. To mitigate the effects of large parallax uncertainties ($\sigma_{\varpi}$), a cut in relative parallax error ($\sigma_{\varpi}/\varpi$) can be applied, assuming that this derives in a biased sample as stated by \cite{GaiaPlx_Luri2018}.
    From now on, the reduced sample verifying $0<\sigma_{\varpi} / \varpi \leq 0.3$ is referred to as the pP30 sample. It contains 1\,394\,075 sources (39.9\% of stars with astrometry). The less restrictive  $0 < \sigma_{\varpi} / \varpi \leq 0.5$ quality cut was used to construct the pP50 sample, which has 2\,064\,351 stars (59.1\% of stars with astrometry).
    Lastly, an analogous sample (hereinafter, named pP30-RV) was created applying the same $0<\sigma_{\varpi} / \varpi \leq 0.3$ cut for those stars having $v_{\rm los}$ from {\it Gaia} DR3. That cut reduced our sample with {\it Gaia} DR3 line-of-sight velocities to 30\,185 stars (94.5\% of them). This sample reaches shorter distances than the previous one (up to $d\approx3$~kpc rather than $d\approx5$-6~kpc) because of the $G_{\rm RVS}=14$~mag limiting magnitude imposed to {\it Gaia} DR3 $v_{\rm los}$. Table~\ref{tab:SubsamplesLen} summarises the sizes of the relevant subsamples described in this section.

    \begin{table}[]
        \centering
        \begin{tabular}{l | l | c}
        Sample & Description & Stars \\
        \hline
            All & A stars with $r\leq19$~mag & 3\,532\,751 \\
            All-DR3 & All + {\it Gaia} DR3 astrometry & 3\,490\,765 \\
            pP50 & All-DR3 + $0<\sigma_{\varpi} / \varpi \leq 0.5$ & 2\,064\,351 \\
            pP30 & All-DR3 + $0<\sigma_{\varpi} / \varpi \leq 0.3$ & 1\,394\,075 \\
            All-RV & All-DR3 + {\it Gaia} DR3 $v_{\rm los}$ & 31\,934 \\
            pP30-RV & pP30 + {\it Gaia} DR3 $v_{\rm los}$ & 30\,185 \\
        \end{tabular}
        \caption{Number of stars of the samples  given in Sect.~\ref{sectionData}.}
        \label{tab:SubsamplesLen}
    \end{table}
    
\begin{figure}
        \centering
        \includegraphics[width=0.9\hsize]{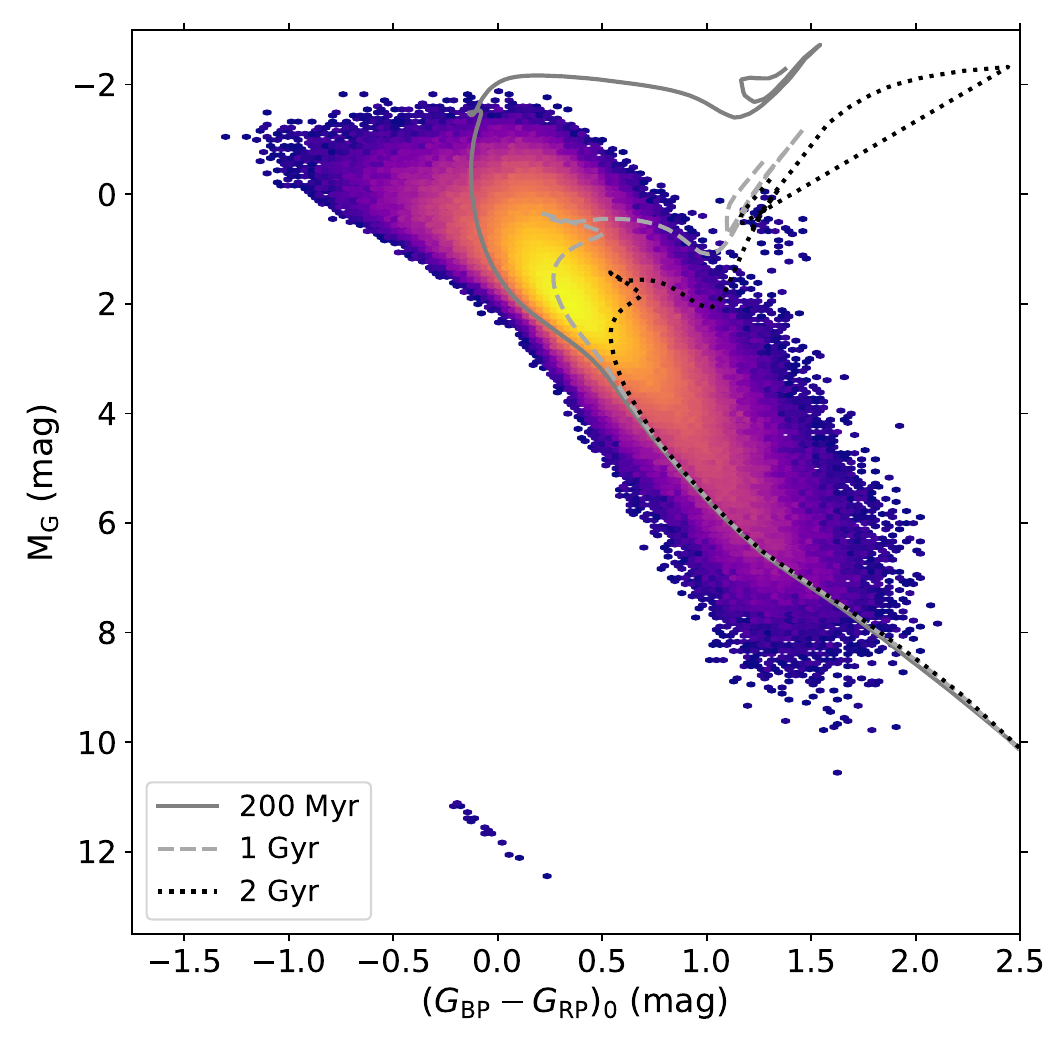}
        \caption{Extinction-corrected {\it Gaia} CMD of the pP30 sample. Three lines indicate the 0.2, 1.0 and 2.0 Gyr isochrones. The colour shows the density in a logarithmic scale where brighter means more stars. Only bins with ten or more stars are plotted.}
        \label{fig:HRcorr}
    \end{figure}

   Figure~\ref{fig:HRcorr} demonstrates again that the level of contamination of our selection method is low by showing the extinction-corrected {\it Gaia} colour-magnitude diagram (CMD) for the pP30 sample together with three isochrones from PARSEC v1.2S \citep{PARSECisoc_Bressan2012, PARSECisoc_Chen2015, PARSECisoc_Marigo2017}\footnote{\url{http://stev.oapd.inaf.it/cgi-bin/cmd}.}. We used reddening values from the \cite{Bayestar_Green2019} dustmap converted to extinctions in our bands of interest (i.e. $G$ and \bprp) using Eqs.~\ref{eq:extinction_transformation1} and \ref{eq:extinction_transformation2} defined in Appendix~\ref{appendixExtinction} together with the usual relations:
    \begin{eqnarray}
        \label{eq:extinction_corrections}
        M_G &=& G -5 {\rm log_{10}}(d) +5 -A_G, \\
        (\bprp)_{\rm 0} &=& (\bprp) - E(\bprp),
    \end{eqnarray}
    where $A_G$ is the extinction in the G band; $(\bprp)$ and $(\bprp)_{\rm 0}$ are the observed and the dereddened {\it Gaia} colours, respectively; $G$ and $M_G$ are the apparent and the absolute magnitude in the G band, respectively; and $d$ must be in parsecs. We obtained $E(B-V)$ reddening values from \cite{Bayestar_Green2019} using the \texttt{dustmaps} python package \citep{DustmapsPython_Green2018} and we transformed them to extinction in the V band according to $A_V = 2.742 E(B-V)$ \citep{SchlaflyFinkbeiner2011, Bayestar_Green2019}\footnote{See also \url{https://dustmaps.readthedocs.io/en/latest/modules.html##module-dustmaps.bayestar}.}. The groups of WD and giants -- shown in Fig.~\ref{fig:HRcorr} at coordinates $(0,12)$ and $(1.25,0.50)$, respectively -- represent such a small fraction of the sample (less than 1\% in total) that they barely affect our results. In addition, detected WDs are located very near the Sun (almost all of them are closer than 0.5~kpc), so they lie in the region where there are not enough statistics to ascertain the reliability of these results.

     \begin{figure}
        \centering
        \includegraphics[width=\hsize]{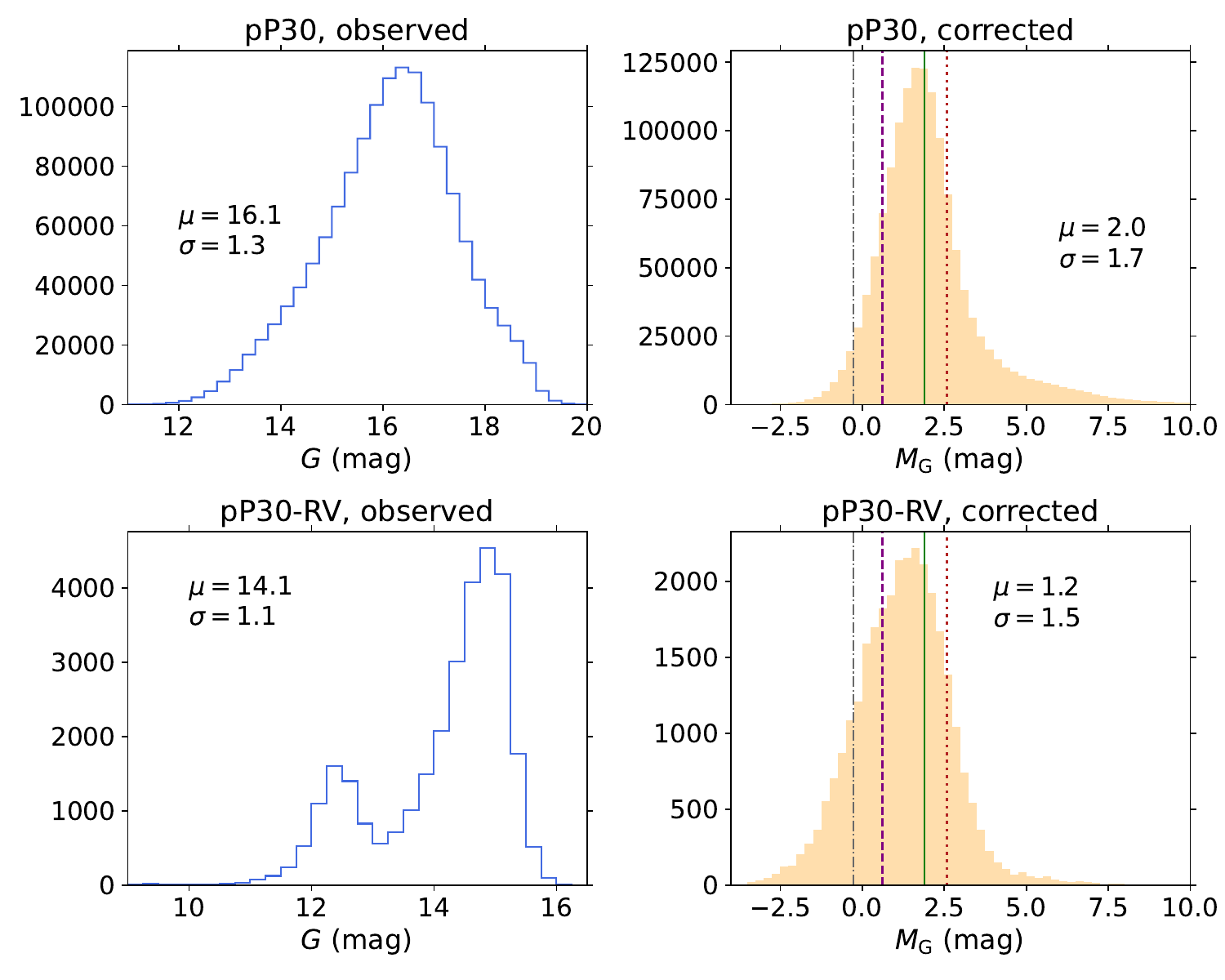}
        \caption{Histograms of apparent magnitude ($G$, left column) and of extinction-corrected absolute magnitude ($M_G$, right column) for the pP30 (top row) and the pP30-RV (bottom row) samples. Vertical lines show absolute $M_G$ magnitudes for B8 (-0.3~mag), A0 (0.6~mag), A5 (1.9~mag), and F0 (2.6~mag) stars from left to right. In all cases, $\mu$ stands for the mean value of the distribution, and $\sigma$ for its standard deviation.}
        \label{fig:MG_hists}
    \end{figure}
    
    In turn, Fig.~\ref{fig:MG_hists} shows the histograms of $G$ and $M_G$ for the pP30 (top) and the pP30-RV (bottom) samples and demonstrates that in both cases the mode is located between the absolute magnitudes expected for A0 and A5 stars (i.e. between 0.6~mag and 1.9~mag)\footnote{All $M_G$ numerical values in this section are derived from the Table~15.7 from \cite{AstropQuantBook} transformed into {\it Gaia} bands according to \url{https://gea.esac.esa.int/archive/documentation/GDR3/Data_processing/chap_cu5pho/cu5pho_sec_photSystem/cu5pho_ssec_photRelations.html} relations.}. The fraction of stars having $M_G$ larger than the value expected for F0 stars (i.e. with $M_G\geq2.6$~mag) is 25.8\% for the pP30 sample and 15.3\% for the pP30-RV one, which confirms again the goodness of the selection. One must take into account that this estimation of the contamination depends on the actual contamination of the selection, but also on the parallax errors and the used distance estimator, as well as on both the applied dustmap and the assumed extinction law. We note that the bottom panel (using the pP30-RV sample) shows that the {\it Gaia}'s selection function for a sample of A-type stars having $v_{\rm los}$ has a clear and artificial bimodality originated by the different methodology applied in the $G_{\rm RVS}\leq12$~mag and the $G_{\rm RVS}>12$~mag regimes \citep{radVelCorrectionDR3_Katz2022}.

\section{Structure}\label{sectionStructureResults}
    This section describes two procedures that shed light on the Milky Way structure through the analysis of stellar densities across the XY Galactic plane. Then, we study the vertical distribution of stars by comparing the sample above and below the plane defined by $b=0\,^{\circ}$.

\subsection{Distribution across the XY plane}\label{sectionXYdistrib}
    We used two different approaches to locate Galactic structures: stellar surface densities and stellar 2D local overdensities. The first one is based on the method described in Sect.~4.1 from \cite{PerArmDetection_Monguio2015}, where the authors define the surface density at each point by extrapolating the observed density in a limited $Z$ range to the whole $Z$ range by assuming a ${\rm sech}^2(Z/h_z)$ vertical density distribution. For our study, we assumed a scale height of $h_z=200$~pc \citep{PerArmDetection_Monguio2015} and ignored the Galactic warp. We also generalised their method splitting the sample into different Galactic longitude bins (of $2\,^{\circ}$) in addition to into distance bins (of 100~pc). The second method involves computing local overdensities with the bivariate kernel density estimation described in Eq.~1 and Appendix B.1 from \cite{Structure_Poggio2021}. We utilised Epanechnikov kernel functions with a local density bandwidth $h_{\rm local}=0.3$~kpc and a mean density bandwidth $h_{\rm mean}=1.5$~kpc applied at intervals of 100~pc in both $X_{\rm Gal}$ and $Y_{\rm Gal}$ coordinates. Although this alternative cannot provide the absolute scale of the densities, it is better at highlighting local overdensities. In fact, they are enhanced by this second method at larger distances (where absolute densities are lower), while remaining partially hidden in the Poisson noise when using surface densities.
    
    \begin{figure*}
        \begin{subfigure}[r]{0.49\hsize}
        \centering
            \includegraphics[width=\hsize]{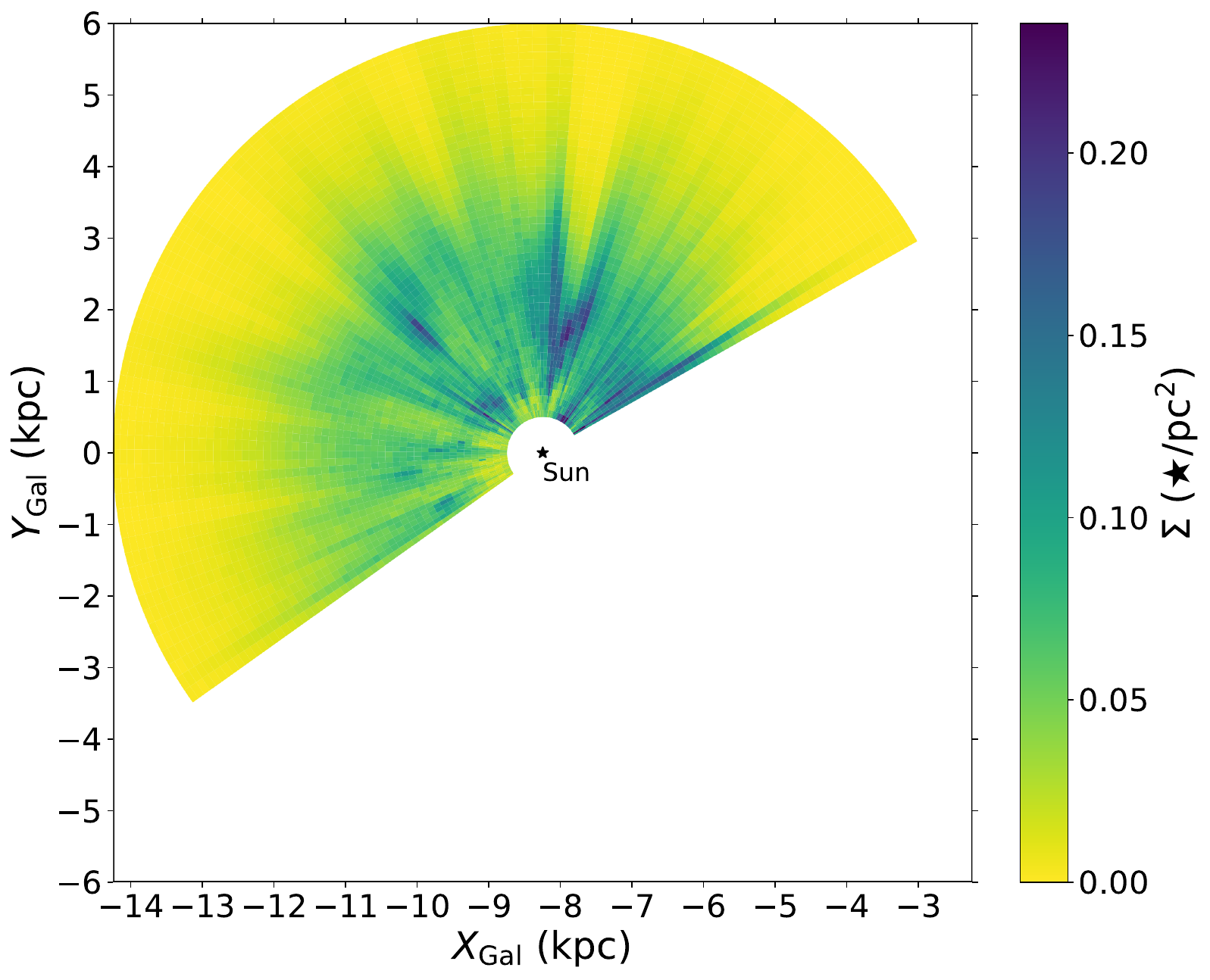}
            \label{subfig:densityMap_pP30}
        \end{subfigure}
        \hfill
        \begin{subfigure}[l]{0.49\hsize}
        \centering
            \includegraphics[width=\hsize]{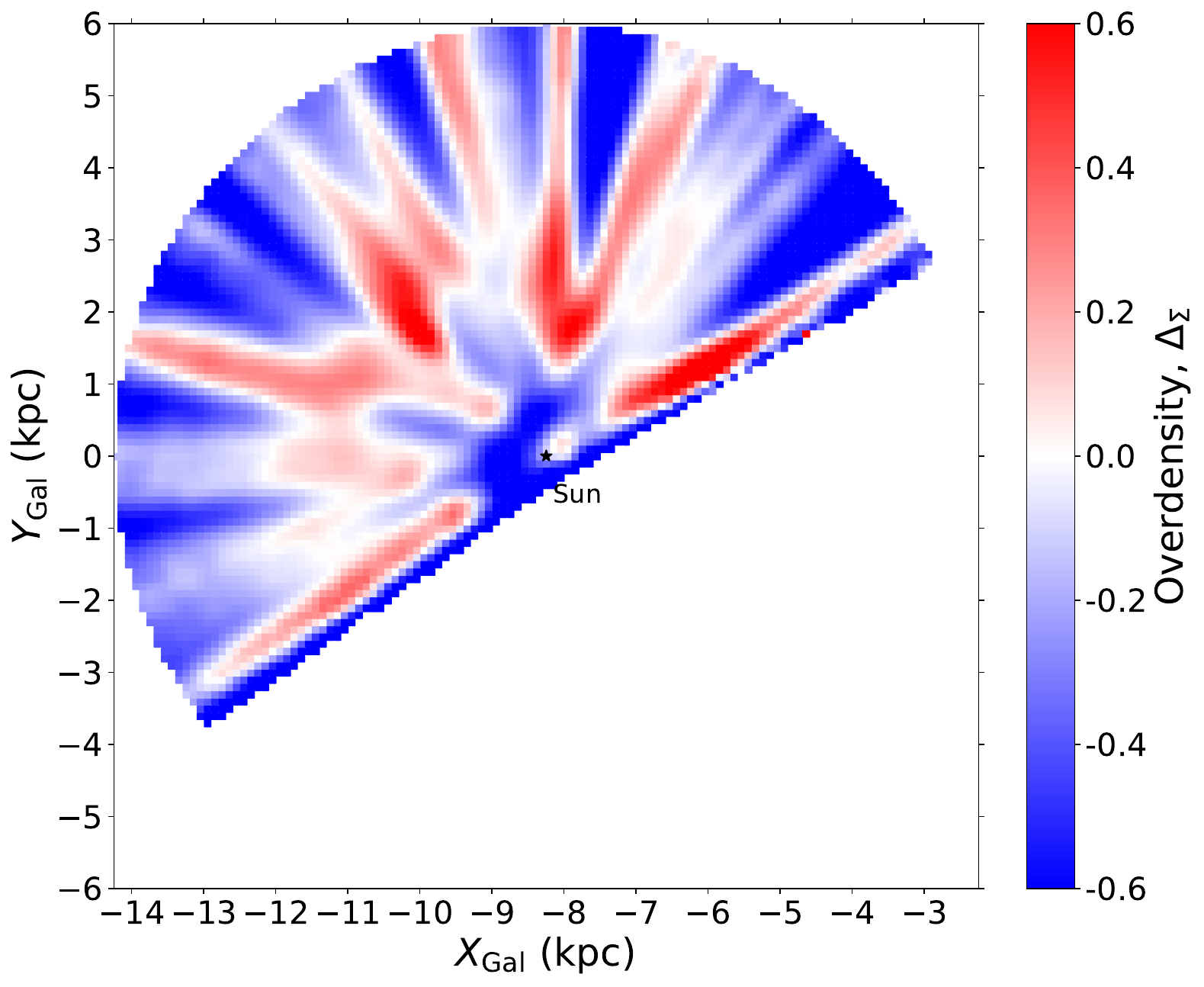}
            \label{subfig:overdensities}
        \end{subfigure}
        \caption{Galactocentric XY maps of the stellar surface density (left) and local overdensities (right) for the pP30 sample. The solar position is indicated with a black star at $X_{\rm Gal}=-R_{\odot}$ and the GC is at the origin, beyond the right-hand edge of each plot.}
        \label{fig:structureXYmaps}
    \end{figure*}

    The stellar surface density map in the XY Galactic plane resulting from applying the first methodology to the pP30 sample is show in the left panel of Fig.~\ref{fig:structureXYmaps}. Densities are very uncertain for $d\leq0.5$~kpc as they rely in a very low number of stars owing to the small volume covered by the $b$-limited sample (thus, they are not shown). The right panel in Fig.~\ref{fig:structureXYmaps} shows stellar local overdensities computed with the second method. The map for the uncut sample is almost completely radial, so it also employs the pP30 sample to avoid severe radial artefacts. In the same way as in its left panel, results from the nearest region are not trustworthy because of the small volume covered by the sample; the furthest values are also unreliable because of parallax uncertainties. This method has some edge effects as well, mainly originated at both $l$ limits. They were partially corrected by taking into account which was the real sampled area.

    Radial structures with respect to the Sun in Fig.~\ref{fig:structureXYmaps} can have two different causes. On the one hand, they can be cones of low density caused by foreground extinction, as the light yellow (blue) triangular region starting around $(X_{\rm Gal},Y_{\rm Gal})\approx(-7.75, \, 2.50)$~kpc in the left (right) panel due to extinction in Cygnus. On the other hand, parallax uncertainties translate into uncertain stellar distances, causing blurring in the radial direction. This second effect elongates density structures, which appear distorted even if they are real. Dealing with these two artefacts is not straightforward. A 3D extinction map is required to estimate the consequences of the former (see Appendix~\ref{appendixCompleteness}), whereas the latter is already mitigated by restricting the sample to high-quality parallaxes.
    
    The resemblance between the two panels of Fig.~\ref{fig:structureXYmaps} allows us to verify the results from the $0 \leq \sigma_{\varpi}/\varpi \leq 0.3$ cut using two different techniques. Their most evident features are (a) the high density spot visible at $(X_{\rm Gal},Y_{\rm Gal})\!\approx\!(-10.00, \, 1.75)$~kpc and (b) the `V' shape starting at $(X_{\rm Gal},Y_{\rm Gal})\!\approx\!(-8, \, 1)$~kpc, which are associated with the Perseus arm and the Cygnus region, respectively. The former may also be associated with a minor peak centred at $(X_{\rm Gal},Y_{\rm Gal})\approx(-10.25, \, 2.50)$~kpc, which is blended with the primary one for the local overdensities. They also agree in showing (c) a short overdense band at $(X_{\rm Gal},Y_{\rm Gal})\approx(-9.00, \, 0.75)$~kpc, very near to the Sun ($d\approx1$~kpc) in the second quadrant; (d) a low-density elongated area centred at $(X_{\rm Gal},Y_{\rm Gal})\approx(-10.0, \, 0.5)$~kpc; and (e) three small peaks relatively close to the AC. In particular, two of them are on both sides of the $Y=0$ line, at $X_{\rm Gal}\approx-9.75$~kpc (only barely visible in the right panel) and at $X_{\rm Gal}\!\approx\!-10.25$~kpc, respectively; the third one is located at $(X_{\rm Gal},Y_{\rm Gal})\approx(-9.50, \, -0.75)$~kpc. In addition, the surface density map shows (f) the beginning of a high-density section at the lower-$l$ end of the sample (i.e. at the right radial cut of this XY map) that extends up to $d\approx3$~kpc. This (f) structure, together with the third peak in (e) are so close to the sample edge, that in the right panel, it is not possible to guarantee whether they are real structures or not. All these (a)-(f) features are also visible when using a cleaner sample with $0 \leq \sigma_{\varpi}/\varpi \leq 0.15$.

\subsection{Vertical distribution}\label{sectionVerticalDistrib}
    This section is devoted to study the vertical stellar distribution. We compared the southern ($b<0\,^{\circ}$) and the northern ($b>0\,^{\circ}$) parts of the Galactic disc by splitting the pP30 sample into these two bins. Then, we computed their stellar surface densities as explained in Sect.~\ref{sectionXYdistrib}. The resulting surface densities are named respectively $\Sigma_{\rm {b<0}}$ and $\Sigma_{\rm {b>0}}$.
    
\begin{figure}
        \centering
        \includegraphics[width=\hsize]{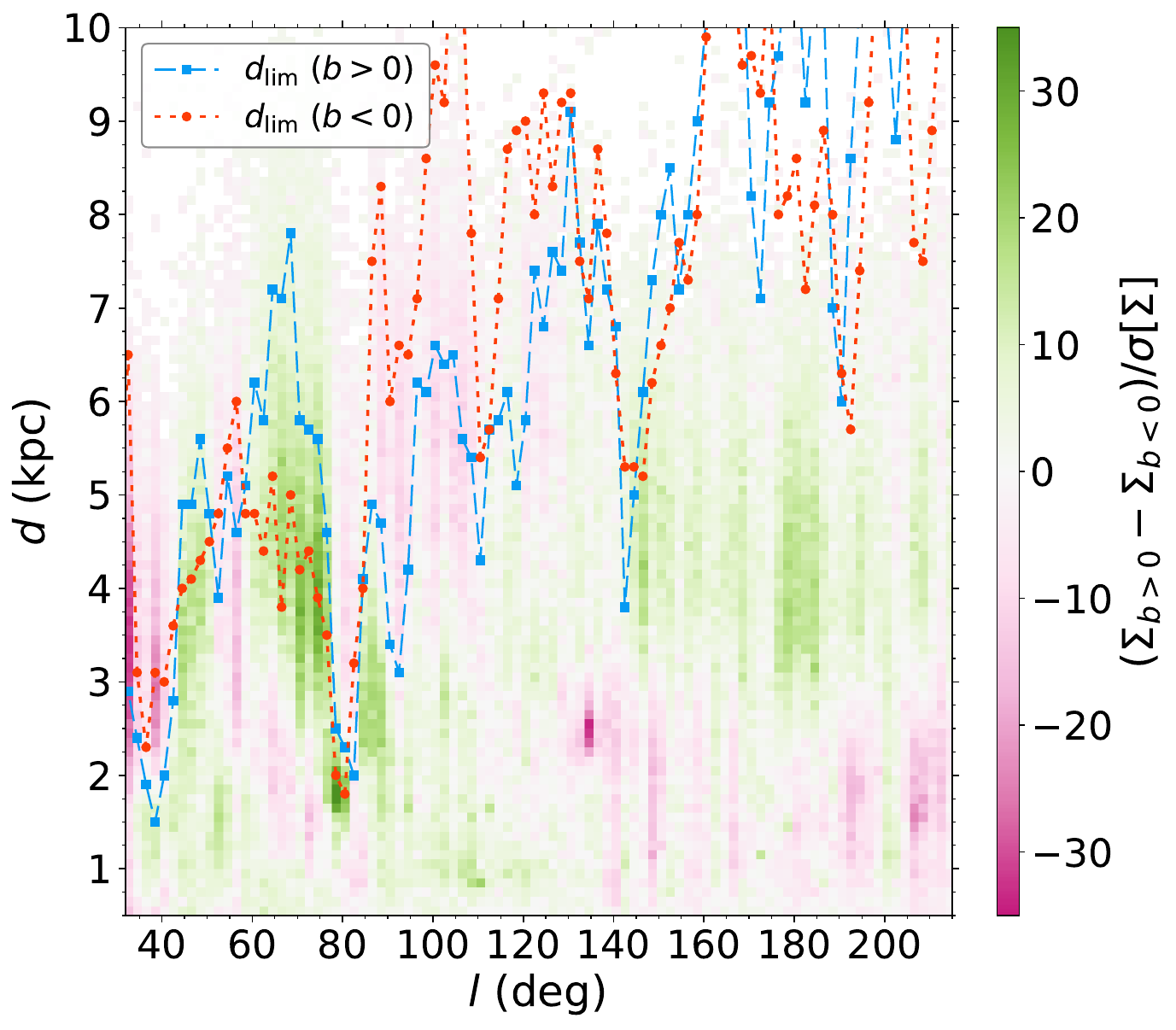}
        \caption{Heliocentric distance versus Galactic longitude map of the difference between the surface densities for $b>0\,^{\circ}$ and $b<0\,^{\circ}$ relative to the uncertainty of the density of the full pP30 sample. Positive values mean larger densities in the northern bin. Estimations of completeness limits for the $b<0\,^{\circ}$ ($b>0\,^{\circ}$) subsample are shown as orange circles (blue squares) connected by a dotted (dashed) line that guides the eye.}
        \label{fig:SupDens_bNPdiff}
    \end{figure}

    In Fig.~\ref{fig:SupDens_bNPdiff}, their difference is normalised in relation to the uncertainty of the surface density of the full pP30 sample, that is: with respect to $\sigma[\Sigma]$, where $\Sigma$ denotes the stellar surface density (left panel of Fig.~\ref{fig:structureXYmaps}). In turn, this uncertainty was estimated by propagating a Poisson error for the number of stars inside each bin \citep[see Eq.~3 from][]{PerArmDetection_Monguio2015}. The employment of this relative deviation allows us to highlight variations that are statistically representative in favour of those which are smaller than uncertainty fluctuations or of the same order. At the same time, $\sigma[\Sigma]$ introduces only small deviations to the general trend of $\Sigma_{b>0}-\Sigma_{b<0}$.
    
    Figure~\ref{fig:SupDens_bNPdiff} reveals structures with discrepancies well beyond the $5\sigma$ level that cannot be explained by simple statistical fluctuations of the stellar density along the vertical direction. So, they originate from actual differences in the observed stellar distribution. However, this may not necessarily correspond with differences in the real Galactic stellar distribution. Once again, extinction may affect the results. For these reason, we include an estimation of the distance ($d_{\rm lim}$) that can be reached with each subsample, restricted to $r\leq19$~mag. These rough estimations of completeness limits, which take extinction into account, are computed as described in Appendix~\ref{appendixCompleteness}.
    
    Despite the fact that they can only be treated qualitatively, a more in-depth analysis of the $d_{\rm lim}$ estimations of completeness limits provides two important outcomes. Firstly, it enables the detection of differences in extinction between the northern and the southern disc. Secondly -- and this aspect is highly correlated with the previous one -- it provides a diagnostic to decide whether a feature in Fig.~\ref{fig:SupDens_bNPdiff} is more probable to be a real asymmetry, or an artefact caused by detection limits due to differences in absorption between both subsamples. The following detailed, step-by-step reasoning clarifies this statement.

    Consider for instance the positive (green) patch at $l\approx70\,^{\circ}$ beyond $d\approx4$~kpc in Fig.~\ref{fig:SupDens_bNPdiff}. Towards this direction, $d_{\rm lim}(b>0) > d_{\rm lim}(b<0)$, which implies that the line of sight can reach deeper distances for $b>0\,^{\circ}$. Therefore, the extinction is higher for $b<0\,^{\circ}$ on average and thus, the number of stars included in a magnitude-limited catalogue would be smaller for $b<0\,^{\circ}$ (where stars are more faded) than for $b>0\,^{\circ}$. This directly results in the observed $\Sigma_{b>0}$ being larger than $\Sigma_{b<0}$ even if the real distribution of stars was completely homogeneous in this location. So, a positive value of the quantity $(\Sigma_{b>0}-\Sigma_{b<0})/\sigma[\Sigma]$ in this context may be caused by extinction biases, at least partially, rather than arising entirely from a real vertical peculiarity in the Galactic disc densities. An analogous argument proves that the opposite stands for $d_{\rm lim}(b>0) \,<\,d_{\rm lim}(b<0)$, as around $l\approx100\,^{\circ}$ or $l\approx120\,^{\circ}$.
    
    For the same reason, the sign of $(\Sigma_{b>0}-\Sigma_{b<0})/\sigma[\Sigma]$ may indicate a real density difference between both subsamples when $d_{\rm lim}(b>0) \approx d_{\rm lim}(b<0)$. For instance, this is the case of the positive (green) spot at $(l,d)\approx(80\,^{\circ},\,1.75\ \rm{kpc})$ or the negative (pink) one at $(l,d)\approx(135\,^{\circ},\,2.5\ \rm{kpc})$. The former indicates that the Cygnus region is predominantly in the northern disc \citep[as seen in the upper panel of Fig.~1 from][]{OBCyg_Quintana2021}.  In contrast, the second example might suggest that the Perseus arm diverts southwards or that there is some other kind of stellar overdensity below the Galactic plane in this direction.
    A detailed analysis of the completeness of each half of the sample is needed to determine up to which distance these densities are truly reliable at each Galactic longitude.
    
\begin{figure}
        \centering
        \includegraphics[width=\hsize]{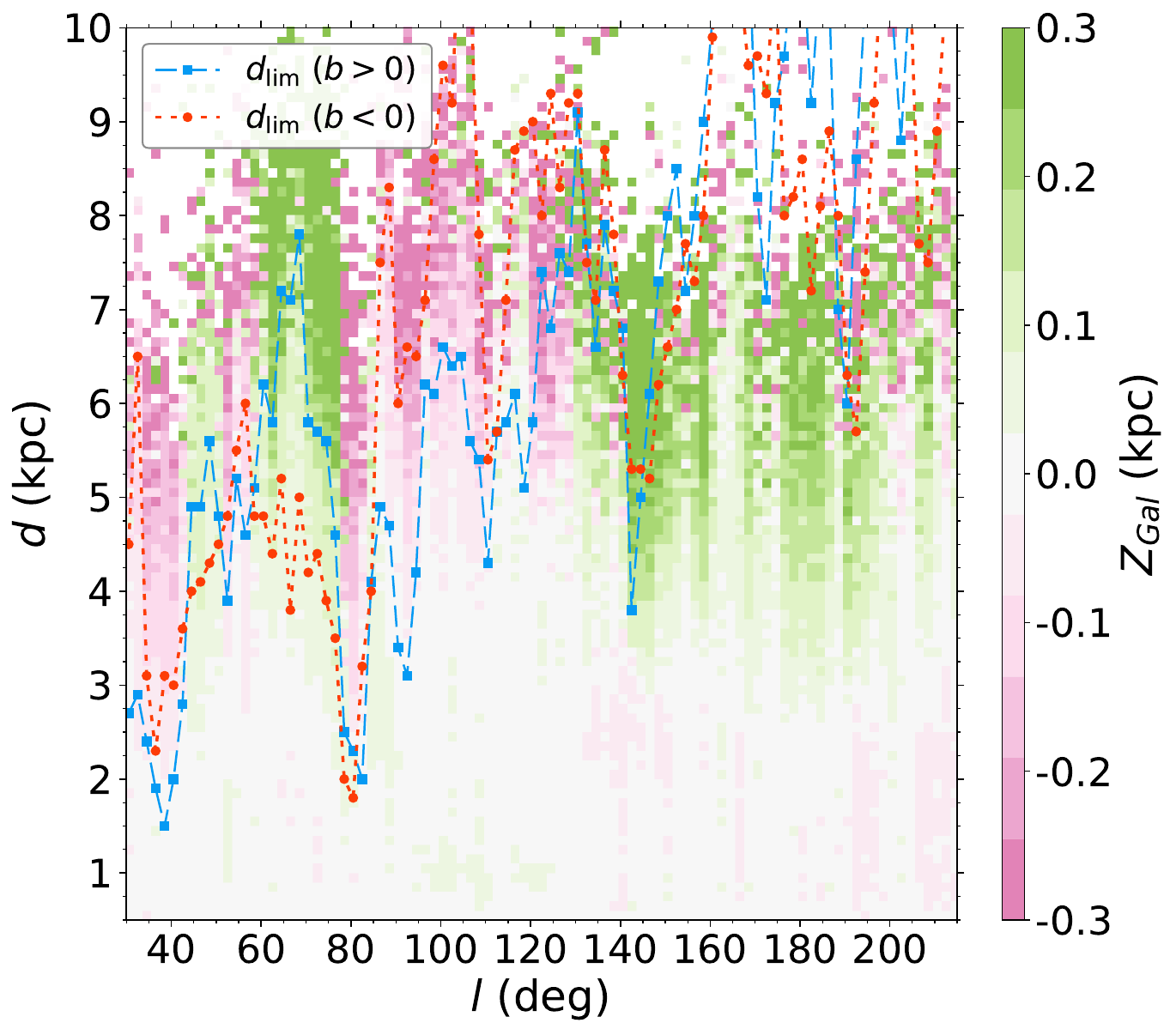}
        \caption{Heliocentric distance versus Galactic longitude map of the distribution of median values of $Z_{\rm Gal}$ for the pP30 sample. Positive values mean that stars are mostly at positive $Z_{\rm Gal}$. Estimations of completeness limits for the $b<0\,^{\circ}$ ($b>0\,^{\circ}$) subsample are shown as orange circles (blue squares) connected by a dotted (dashed) line that guides the eye.}
        \label{fig:Zmedian}
    \end{figure}

    In Fig.~\ref{fig:Zmedian}, we plot the median values of the $Z_{\rm Gal}$ coordinate for the pP30 sample. The sample is binned in $l$ and $d$ employing the same grid as before (i.e. $2\,^{\circ}$ width in $l$ and 100~pc long in $d$). We note that the range in $Z_{\rm Gal}$ of the closest bins to the Sun is extremely limited owing to the sample restriction in $b$ (reaching only $|Z_{\rm Gal}|=87$~pc at $d=1$~kpc). Therefore, their medians will always be systematically near zero.
    This figure has features equivalent to those in Fig.~\ref{fig:SupDens_bNPdiff}, though they tend to start about 1-2~kpc later for median $Z_{\rm Gal}$ than for $(\Sigma_{b>0}-\Sigma_{b<0})/\sigma[\Sigma]$. The main examples of this are the positive (green) values at $60\,^{\circ}\lesssim l\lesssim80\,^{\circ}$ and at $l\gtrsim130\,^{\circ}$, or the negative (pink) region between them. This demonstrates that both approaches are highly correlated.
    
    We note that these two figures are dependent on the value assumed for $Z_{\odot}$. However, with a value of $Z_{\odot}=20.8$~pc \citep{Zsun_Bennett2019}, the change in Fig.~\ref{fig:Zmedian} is a negligible shift. The $\Sigma_{b>0}-\Sigma_{b<0}$ seen in Fig.~\ref{fig:SupDens_bNPdiff} has small variations (smaller than 10 times $\sigma[\Sigma]$ for 90\% of the bins) that do not modify the main structures.
    
    The Galactic warp is a large-scale structure bending the disc, shifting the density upwards in the northern Galactic plane. As seen through HI gas, it has an amplitude (i.e. height at the direction of maximum deviation, $\phi\approx90\,^{\circ}$) between 1.3~kpc \citep{HIwarp_Levine2006} and 1.7~kpc \citep{HIspiralArms_Koo2017} at $R\approx16$~kpc. Cepheids show an amplitude of 1.0~kpc at $R\approx14$~kpc \citep{WarpCepheids_Skowron2020}, while \cite{Warp_Merce2019} find 0.2~kpc amplitude for OB stars and 1.0~kpc for RGB stars at the same Galactocentric radius.
    If there were no extinction, we would expect that median $Z_{\rm Gal}$ values grow according to the previous characterisation.
    As shown in Fig.~\ref{fig:Zmedian}, our sample is highly affected by differences in the extinction above and below the plane. Therefore, no clear conclusions about the warp using only stellar counts can be made at this stage. In the next section, we study the warp through kinematics, which are not so heavily affected by incompleteness.

\section{Kinematics}\label{sectionKinematics}
    This section focuses on our A-type stars large-scale kinematics, in particular, on the $(V_R,\,V_{\phi},\,V_Z)$ velocity maps (Sect.~\ref{subsectionKinematicsWithRV}), on the $v_b^{\rm corr}$ distribution and the Galactic warp (Sect.~\ref{subsectionKinematicsNoRV}), and on the inhomogeneities of the $V_Z$-$V_{\phi}$ plane (Sect.~\ref{subsectionVelSpaceAssym}).

\subsection{6D sample}\label{subsectionKinematicsWithRV}
\begin{figure}
        \centering
        \begin{subfloat}
        \centering
            \includegraphics[width=0.97\hsize]{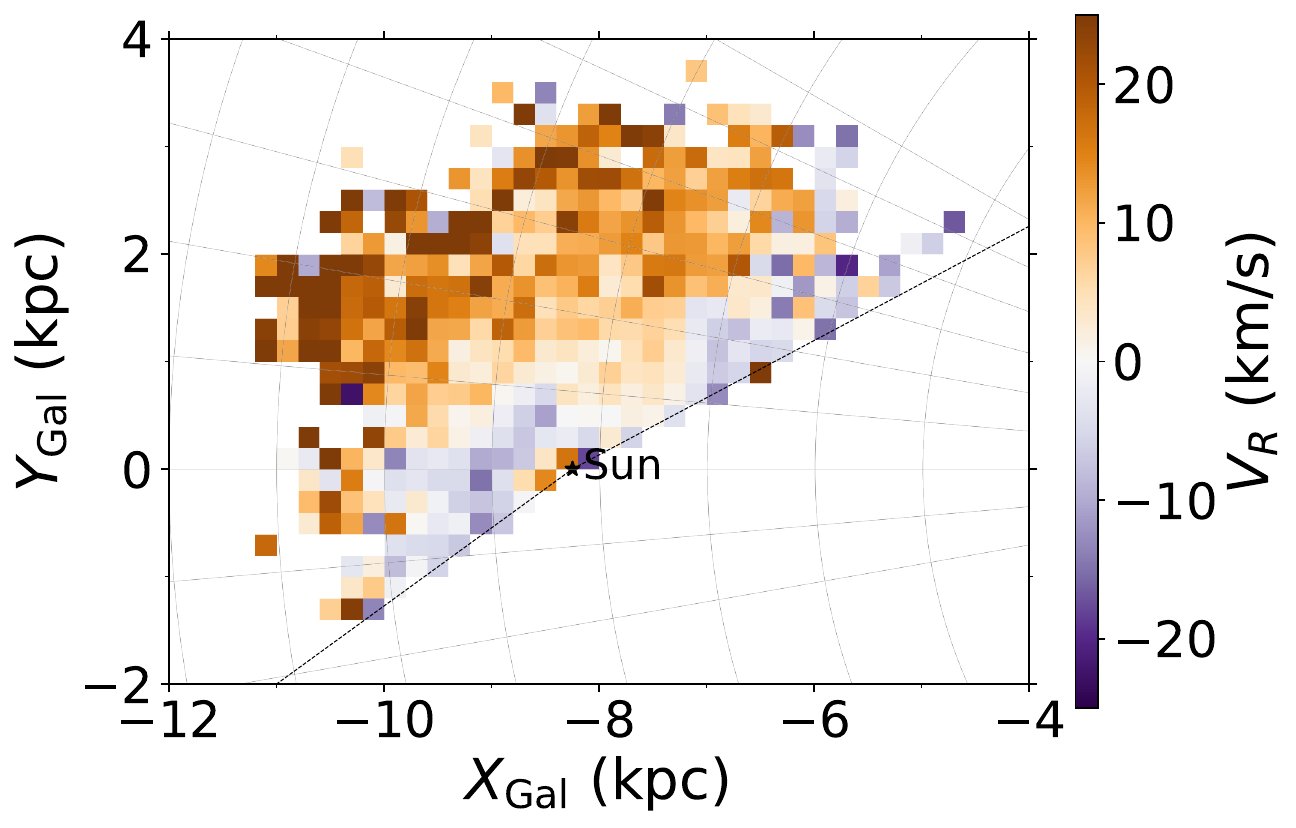}
        \end{subfloat}
        \vfill
        \begin{subfloat}
        \centering
            \includegraphics[width=0.97\hsize]{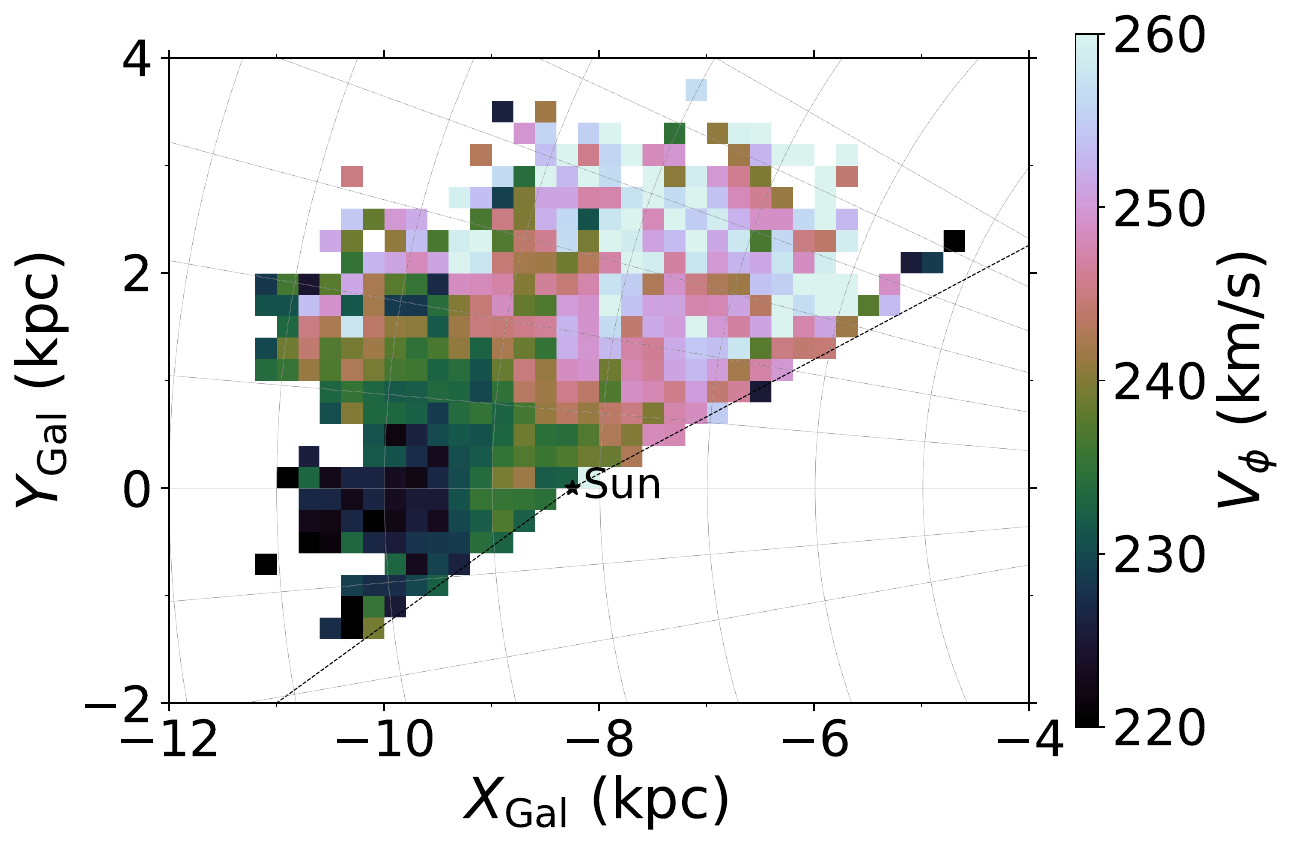}
        \end{subfloat}
        \vfill
        \begin{subfloat}
        \centering
            \includegraphics[width=0.97\hsize]{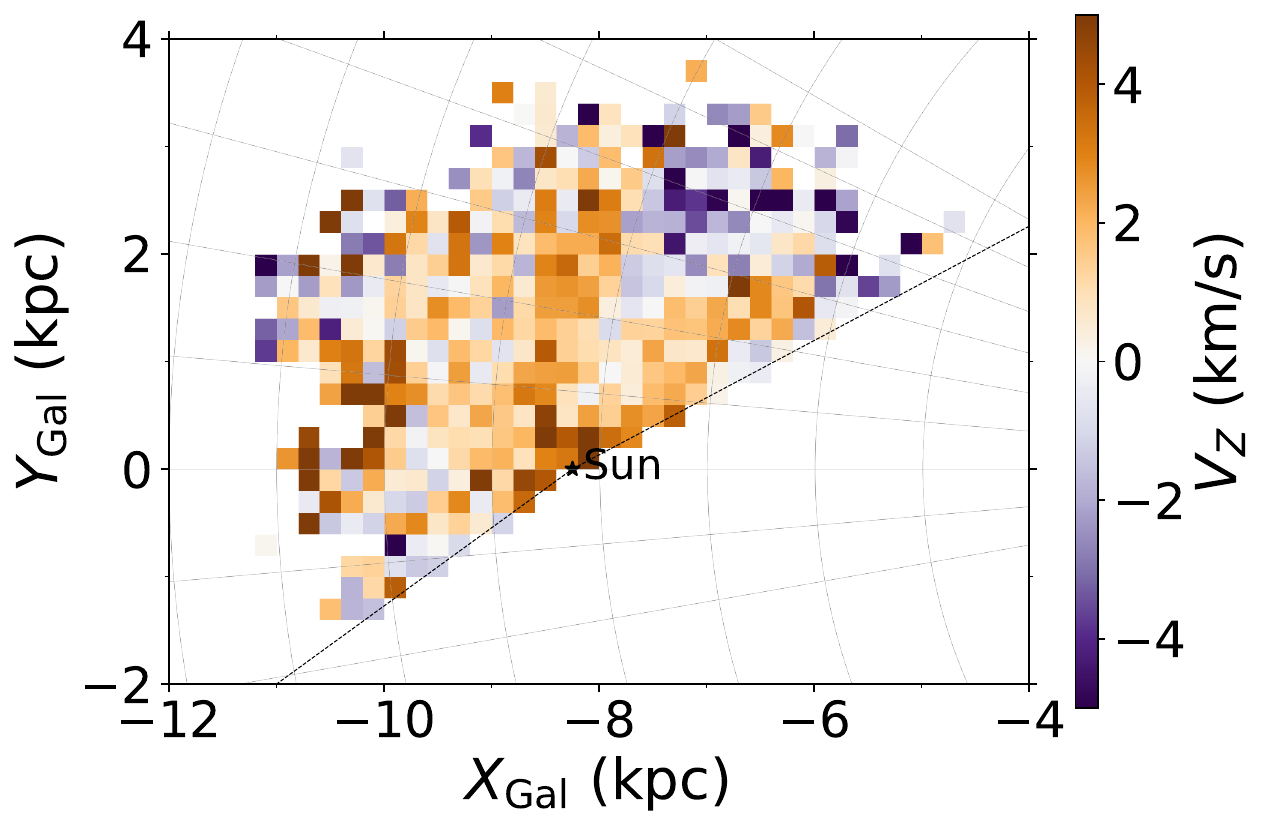}
        \end{subfloat}
        \caption{Velocity maps of the pP30-RV sample in the Galactocentric XY plane. They show medians of radial ($V_R$), azimuthal ($V_{\phi}$), and vertical ($V_Z$) velocities from top to bottom. The bins have 0.2~kpc per side and only those containing ten or more stars are shown. Black lines show the Galactic longitude limits of the sample. Grey lines indicate Galactocentric radii from 5 to 12~kpc and azimuths from $-15\,^{\circ}$ to $30\,^{\circ}$ with $5\,^{\circ}$ steps.}
        \label{fig:VrVphiVz-maps_DR3}
    \end{figure}

    The Galactocentric XY maps for the pP30-RV sample in the three components of cylindrical Galactocentric velocities $V_R$, $V_{\phi}$, and $V_Z$ are shown in Fig.~\ref{fig:VrVphiVz-maps_DR3}. We note that due to the very strict constrains of the $v_{\rm los}$ requirement, the sample barely changes when applying the parallax quality cut (two last rows of Table~\ref{tab:SubsamplesLen}). The main properties of these maps are described below. We only discuss those features extending far from the edges and over a large amount of bins.
    
    In the top panel, median Galactocentric radial velocities are positive (outwards) at $l\approx45$-$90\,^{\circ}$ and at the second quadrant, whereas it presents two negative clumps close to the sample lower borders.
    In the middle panel, the azimuthal velocity map has important variations with considerably low values ($V_{\phi}\lesssim230\,\rm{km\,s^{-1}}$) towards the AC and values above $240\,\rm{km\,s^{-1}}$ in all the first quadrant. When looking at the AC direction, we see a rapidly decreasing trend having a $\sim$~$15\,\rm{km\,s^{-1}}$ drop within a range of $\sim$~$2$~kpc in $R$. This could be related with the asymmetric drift that causes median $V_{\phi}$ to be lower than circular velocities \citep[chapter 4.8.2a from ][]{GalacticDynamics_BT2008}. However, due to the small radial velocity dispersion of our relatively young targets \citep[around $15\,\rm{km\,s^{-1}}$ according to][]{VelocityDispersionFig2_Aumer2009}, this effect is almost independent of $R$ and smaller than $5\,\rm{km\,s^{-1}}$ for our A-type stars \citep[see right panel of Fig. 4 from][]{AsymDrift_Robin2017}. Thus, this effect cannot explain the large peculiarities we find.
    In the bottom panel, the most striking feature is a region with downward motion ($V_Z<0\,\rm{km\,s^{-1}}$) at $l\approx60$-$75\,^{\circ}$, around $(X_{\rm Gal},Y_{\rm Gal})\approx(-7.0,\,2.5)$~kpc. It is surrounded by two elongated regions with upward motion, one at larger $l$ and the other at lower $l$ with respect to it. All this indicates the presence of a kinematic perturbation in the aforementioned region. Considering that at $l\approx60$-$75\,^{\circ}$ nearby extinction is more concentrated below $b=0\,^{\circ}$ than above (see the difference between both completeness distance estimations $d_{\rm lim}$ in Fig.~\ref{fig:SupDens_bNPdiff}, the explanation in Sect.~\ref{sectionVerticalDistrib}, Figs.~3-4 from \citealp{ExtinctionMap_Sale2014} or also Fig.~2 from \citealp{Bayestar_Green2019}), this perturbation may be the signature of a compression breathing mode. Galactic warp effects are expected to start too far \citep[beyond $R\approx10$-$12$~kpc according to][]{Warp_Merce2019, AssymDiscDR3_Drimmel2022} to be observed with this reduced sample.
    
    We expect that some of these detected features are related with perturbations caused by spiral arms or by interactions with the Milky Way satellites (such as the Sagittarius dwarf galaxy). The short distance covered and the lack of part of the first and the third quadrants, as well as the full fourth quadrant, make it difficult to reach any further conclusion in this regard. Acquisition of a larger number of line-of-sight velocities for our A-type stars sample will allow us to clarify detected kinematic patterns and to fully understand them.

\subsection{Without the line-of-sight velocities}\label{subsectionKinematicsNoRV}
    Our IGAPS selection and the recent {\it Gaia} DR3 make these A-type stars kinematic maps  possible up to $r=19$~mag for the first time. In this section we improve the statistics of the previous 3D kinematic analysis and we study $v_b^{\rm corr}$ (without $v_{\rm los}$) for the pP50 sample.

    Appendix~\ref{appendixKinModel} explains the effects of using $v_b^{\rm corr}$ as an estimator for $V_Z$ and the differences between these two magnitudes. In particular, it is worth to notice that as the sample is limited to $30\,^{\circ} \lesssim l\lesssim 215\,^{\circ}$, the main expected discrepancy introduced by this approximation is a compression breathing mode dependent on $l$ that is present in the $v_b^{\rm corr}$ distribution, but not in the $V_Z$ one.
    
\begin{figure}
        \centering
        \includegraphics[width=\hsize]{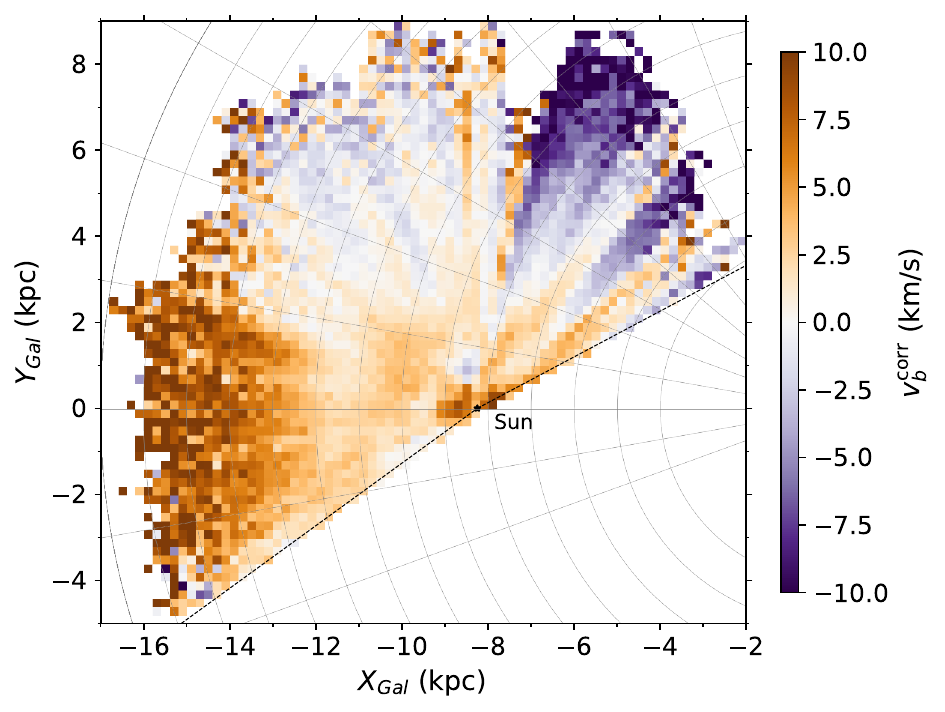}
        \caption{Galactocentric XY map of the corrected velocities in the $b$ direction derived from {\it Gaia} DR3 proper motions for the pP50 sample. Bins have sides equal to 0.2~kpc and are colour-coded by median $v_b^{\rm corr}$. Only those bins with ten or more stars are shown. Black diagonal lines show the Galactic longitude limits of the sample. Grey lines indicate Galactocentric radii from 4 to 16~kpc and azimuths from $-20\,^{\circ}$ to $70\,^{\circ}$ with $10\,^{\circ}$ steps.}
        \label{fig:XY_vbcorr_EDR3}
    \end{figure}

    The map of median $v_b^{\rm corr}$ across the Galactic XY plane is shown in Fig.~\ref{fig:XY_vbcorr_EDR3}. It has two prominent features. The first one is a very negative $v_b^{\rm corr}$ region at $(l,\,d)\approx(60$-$75\,^{\circ},\,6$-$7\ \rm{kpc})$ -- or (equivalently) around $(X_{\rm Gal},Y_{\rm Gal})\approx(-6,\,6)$~kpc --, which coincides with the prolongation of the $V_Z<0\,\rm{km\,s^{-1}}$ peculiarity highlighted in the previous section with the 6D sample (bottom panel of Fig.~\ref{fig:VrVphiVz-maps_DR3}). This feature might be artificially enhanced in Fig.~\ref{fig:XY_vbcorr_EDR3} by the use of $v_b^{\rm corr}$ as a proxy of $V_Z$ (see Appendix~\ref{appendixKinModel}).
    The second feature is a positive $v_b^{\rm corr}$ band towards $160\,^{\circ}\lesssim l\lesssim 210\,^{\circ}$ beyond $R\!\approx\!12$~kpc (left-most edge of the sample). It demonstrates the existence of the Galactic warp signature as a large-scale structure located approximately around the AC at large Galactocentric radii ($R\!\gtrsim\!12$~kpc) that is moving upwards. No important differences between $v_b^{\rm corr}$ and $V_Z$ are expected in this region (see Fig.~\ref{subfig:1725l1875_Model2}).

\begin{figure}
        \centering
        \includegraphics[width=\hsize]{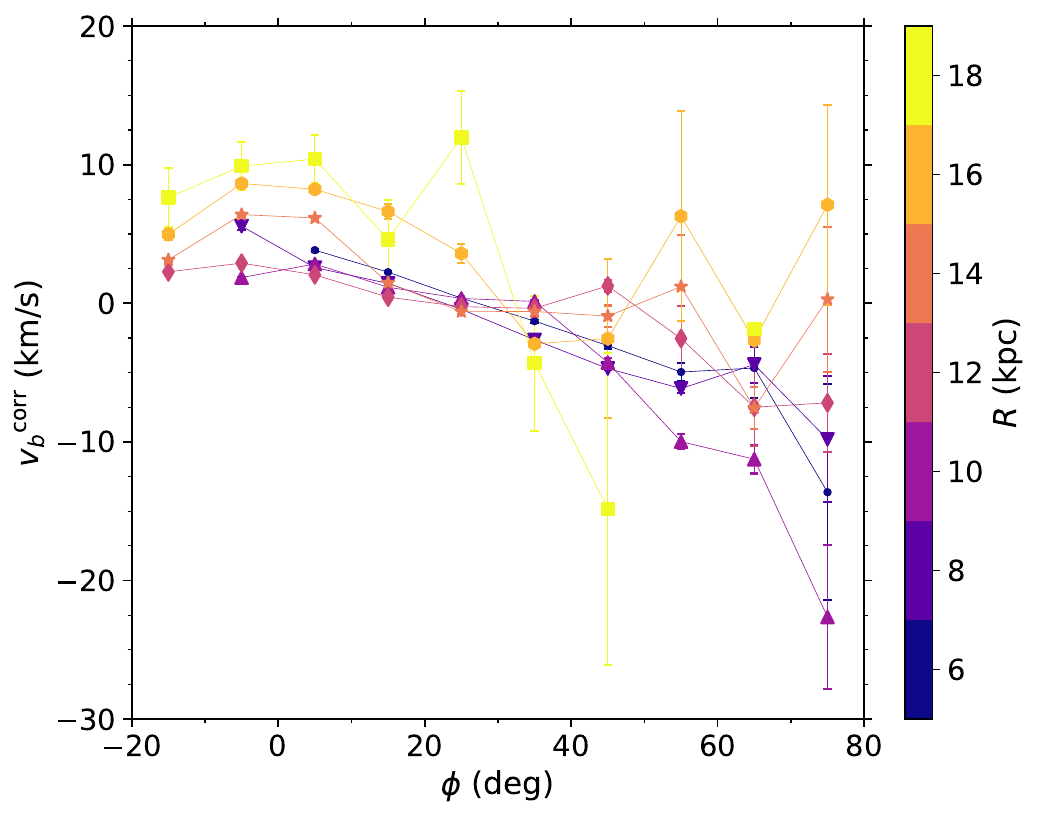}
        \caption{Median velocity in $b$ direction, corrected for the solar motion, versus the Galactocentric azimuthal coordinate. The pP50 sample sources are binned in $R$ every 2~kpc. Error bars show standard deviations divided by the square root of the number of stars in each bin.}
        \label{fig:warp_vbcorrVSphi_EDR3}
    \end{figure}

    A Galactic warp causes stellar orbits to have a specific vertical waving. In particular, for a symmetric warp formed by tilted flat circular orbits \citep[see Fig.~5.1 from][]{Thesis_Hoda2015} its maximum and minimum vertical velocities are reached towards the line-of-nodes \citep[this would not necessarily be the same for different warp geometries, see for instance][]{Warp_Merce2019}. Furthermore, the amplitude of this oscillating motion is expected to grow with Galactocentric radius $R$, as so does the maximum vertical coordinate reached. Figure~\ref{fig:warp_vbcorrVSphi_EDR3} confirms this predictions by showing median $v_b^{\rm corr}$ as a function of $\phi$ for different bins in Galactocentric radius, reaching up to $v_b^{\rm corr}\approx6$-$7 \,\rm{km\,s^{-1}}$ at $R\approx14$~kpc.
    
\begin{figure}
        \centering
        \begin{subfigure}[b]{0.49\hsize}
        \centering
            \includegraphics[width=\hsize]{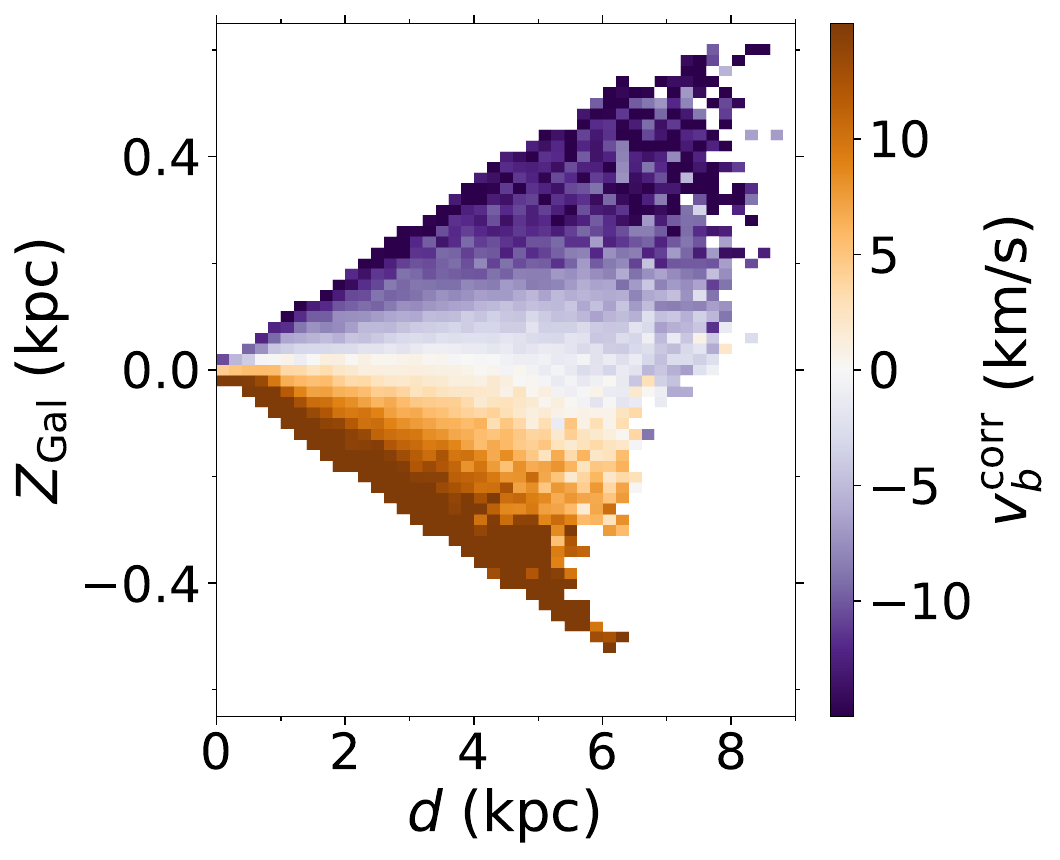}
            \caption{$60 \,^{\circ} \leq l \leq 75 \,^{\circ}$}
            \label{subfig:60l75_EDR3}
        \end{subfigure}
        \hfill
        \begin{subfigure}[b]{0.49\hsize}
        \centering
            \includegraphics[width=\hsize]{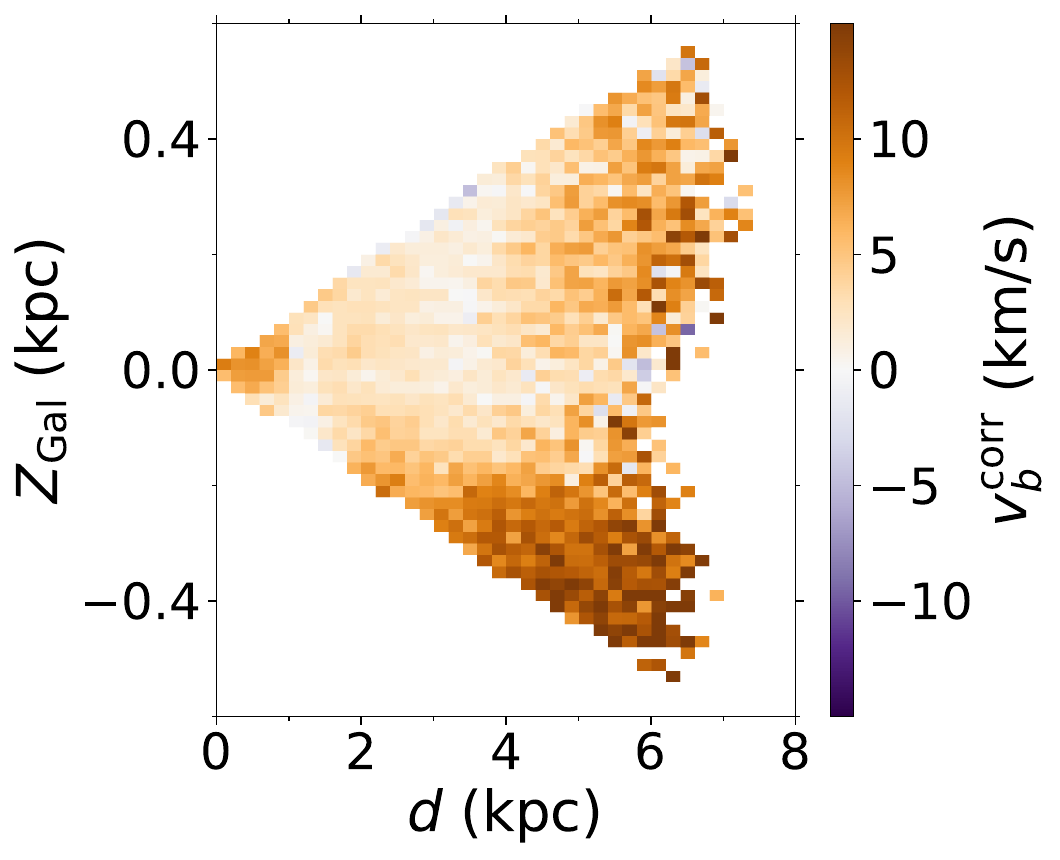}
            \caption{$172.5 \,^{\circ} \leq l \leq 187.5 \,^{\circ}$}
            \label{subfig:1725l1875_EDR3}
        \end{subfigure}
        \caption{Vertical Galactic coordinate as function of heliocentric distance, colour-coded by the median $v_b^{\rm corr}$ for the pP50 sample. Panel (a) displays the slice between $60\,^{\circ}\leq l\leq75\,^{\circ}$, whereas panel (b) uses stars with $172.5\,^{\circ}\leq l\leq187.5\,^{\circ}$. In both cases, the size of the bins is 0.2~kpc in $d$ and 20~pc in $Z_{\rm Gal}$. Bins with fewer than ten stars are not shown.}
        \label{fig:Zd_vbcorrVZ_EDR3}
    \end{figure}

    The smallest $v_b^{\rm corr}$ values seen in Fig.~\ref{fig:warp_vbcorrVSphi_EDR3} beyond $\phi\approx50\,^{\circ}$ correspond with the negative $v_b^{\rm corr}$ feature found in Fig.~\ref{fig:XY_vbcorr_EDR3} around $(X_{\rm Gal},\,Y_{\rm Gal})\approx(-6,\,6)$~kpc. To revisit this region and analyse it in depth, two slices in $l$ were selected from the pP50 sample (as it was done for the Appendix~\ref{appendixKinModel} model): namely $60\,^{\circ}\leq l\leq 75\,^{\circ}$ and $172.5\,^{\circ}\leq l\leq 187.5\,^{\circ}$ (the second one as a reference to compare a direction where there are almost no differences between $v_b^{\rm corr}$ and $V_Z$). Figure~\ref{fig:Zd_vbcorrVZ_EDR3} shows how $v_b^{\rm corr}$ is distributed in the $Z_{\rm Gal}$-$d$ plane for each of them. Figure~\ref{subfig:60l75_EDR3} presents a noticeable compression breathing mode that is very similar in amplitude to the one modelled in Fig.~\ref{subfig:60l75_Model2} (note that these two panels have almost the same colour range). A relevant difference between the model and the data is that the former displays a straight division between positive and negative velocities (thanks to its high degree of symmetry), while the band satisfying $v_b^{\rm corr}\approx0\,\rm{km\,s^{-1}}$ widens and seems to bend towards negative $Z_{\rm Gal}$ at $d\gtrsim3.5$~kpc for the pP50 sample. By contrast, Fig.~\ref{subfig:1725l1875_EDR3} exhibits globally positive medians of $v_b^{\rm corr}$ caused by the Galactic warp. Furthermore, it also shows a compression breathing mode, although much weaker than the previous one. In this case, neither the warp signature nor the observed breathing mode cannot be attributed to the differences between $v_b^{\rm corr}$ and ${V_Z}$.
    
\subsection{Velocity space asymmetries}\label{subsectionVelSpaceAssym}
    This section further exploits the strengths of A-type stars as Galactic-scale tracers by analysing the velocity space asymmetries. We study the distribution in the $V_Z$-$V_{\phi}$ plane, which displays several overdensities whose shape and size (and even their number in some cases) depend on $R$ as well as on $Z_{\rm Gal}$. Similar plots are available in Fig.~13 of \cite{AC_Teresa2021} for sources without any selection on stellar types \citep[see also][where  these perturbations are analysed based on a sample of intrinsically blue stars]{Disturbances_McMillan2022}.
    
\begin{figure*}[!]
        \centering
        \includegraphics[width=\hsize]{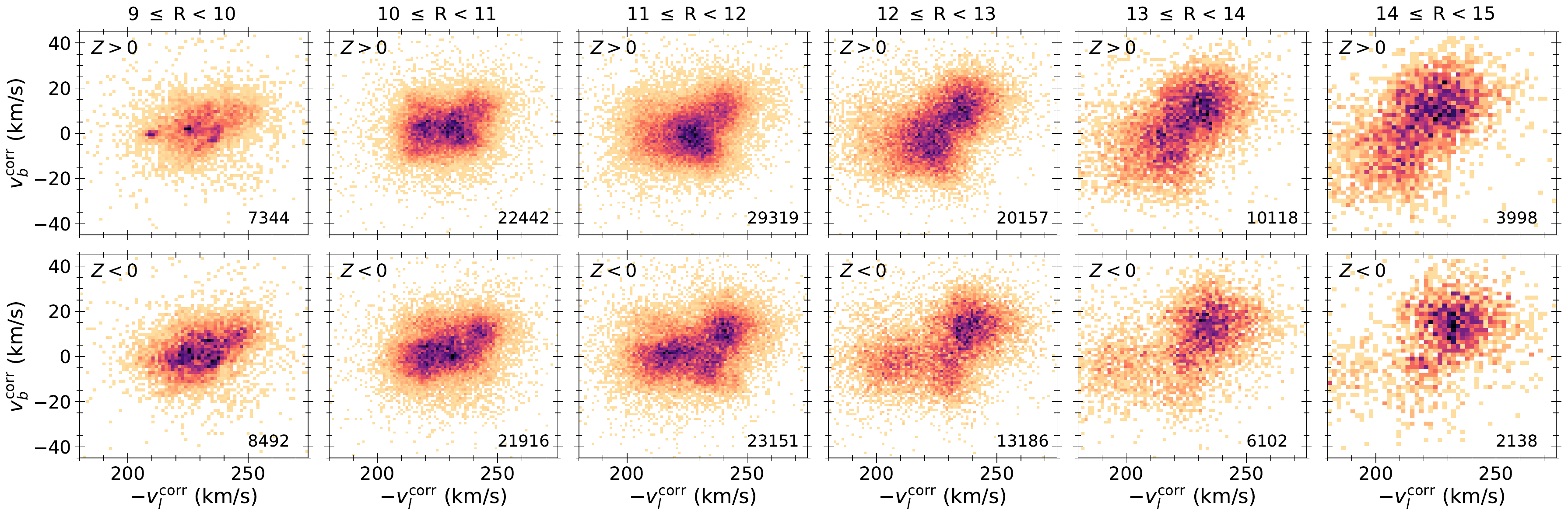}
        \caption{Density in the $V_Z\equiv v_b^{\rm corr}$ vs $V_{\phi}\equiv -v_l^{\rm corr}$ velocity map for several Galactocentric radial bins for the pP50 sample restricted around the AC ($170\,^{\circ}\leq l\leq190\,^{\circ}$). The two rows distinguish $Z_{\rm Gal}>0$ (above) from $Z_{\rm Gal}<0$ (below). All panels share both axes. The colour indicates stellar counts normalised to the densest bin of each panel in the sense that darker means denser. Bin sizes have been adapted to the number of stars of each panel (numbers shown at bottom right corners).}
        \label{fig:VZvsVphi_pP50}
    \end{figure*}

    Since the current study requires to use narrow Galactocentric radial bins, the pP50 sample is used again to improve the statistics.
    We use the same approach as in \cite{AC_Teresa2021}, with the approximations $V_Z \equiv v_b^{\rm corr}$ and $V_{\phi} \equiv -v_l^{\rm corr}$ for a sample of sources restricted to $170\,^{\circ}\leq l\leq190\,^{\circ}$. The resulting distribution in the $v_b^{\rm corr}$ vs $-v_l^{\rm corr}$ plane for several bins in Galactocentric radius is shown in Fig.~\ref{fig:VZvsVphi_pP50}, where the two rows compare stars having $Z_{\rm Gal}>0$ (upper row) with those having $Z_{\rm Gal}<0$ (bottom row).
    
    The most compact peaks seen in the upper left panel of this figure at $(v_b^{\rm corr},-v_l^{\rm corr})\approx(-0.3,\,210)\,\rm{km\,s^{-1}}$ and $(v_b^{\rm corr},-v_l^{\rm corr})\approx(1.9,\,225)\,\rm{km\,s^{-1}}$ are associated with overdensities in sky coordinates; actually, they are the signature of open clusters NGC 2099 and NGC 1912, respectively. On the other hand, Fig.~\ref{fig:VZvsVphi_pP50} displays clear velocity space inhomogeneities, as well as asymmetries between above and below the Galactic plane. By visual inspection of panels beyond $R=11$~kpc, two main peaks are seen for $Z_{\rm Gal}>0$, whereas stars at $Z_{\rm Gal}<0$ are grouped in three major overdensities. All of them move towards smaller rotational velocities and slightly smaller vertical velocities when increasing $R$. The origin of this substructure is still controversial, although it could be related with the passage of a satellite as the Sagittarius dwarf galaxy \citep{Disturbances_McMillan2022}.
    
    Since $v_{\rm los}$ information is very relevant to disentangle the origin of these inhomogeneities, we tried to reproduce these results using $V_{\phi}$ and $V_Z$ velocities for stars with {\it Gaia} DR3 line-of-sight velocities. Nevertheless, they are so few that no kinematic substructure can be recovered. To recover, at least partially, the large statistics of the original A-type stars sample and complete these analysis using the 3D velocity space, $v_{\rm los}$ measurements from WEAVE will be crucial. In contrast with {\it Gaia}, whose line-of-sight velocities methods (measuring the near infrared calcium triplet region) are more suitable for colder stars, WEAVE will have a survey specially dedicated to A-type stars \citep{WEAVE_Jin2022} which will definitively help to fill this gap.

\section{Discussion}\label{sectionDiscussion}
    This section is devoted to discuss the major findings of the work, to contrast density and kinematics of A-type stars with different tracers and to compare our results with the literature. It follows the same order as previous sections, starting by density structures and then discussing kinematics.

    Overdense structures highlighted in Fig.~\ref{fig:structureXYmaps} are likely associated with: (a) the Perseus arm; (b), (c), and (e) different parts of the Local arm; and (f) the largest-$l$ component of the Sagittarius-Carina arm. This is in agreement with features found by \cite{Structure_Poggio2021} in their Fig.~1. In contrast, our structure (c) can also be related with the Cepheus spur defined by OB stars in the left panel of Fig.~5 from \cite{CepSpur_Pantaleoni2021}. We have checked that the spiral arms models of \cite{SpiralArms_Xu2018} and \cite{SpiralArms_Reid2019} overlap with the overdensities found in Perseus, Local and Sagittarius-Carina arms.
    However, the observed distribution favours an scenario with clumpy structures over continuous spiral arms. The low density region (d) arises naturally from the inter-arm region between the Perseus and the Local arms based on the \cite{Structure_Poggio2021} approach. On the contrary, \cite{OCvoid_CantatGaudin2020} highlights, in the same coordinates, a lack of open clusters that splits their more tightly wound model of the Perseus arm (see the right panel of their Fig.~11). Regardless of which pitch angle is the correct one, there is an `empty' volume at this position. At least, it is depleted of masers \citep{SpiralArms_Reid2019}, upper main sequence stars \citep{Structure_Poggio2021}, A-type stars (this work) and open clusters (\citealp{OCvoid_CantatGaudin2020} and references in their Sect. 5.1, \citealp{OCgdr3_Hunt2023}). Furthermore, it is also recovered by \cite{AssymDiscDR3_Drimmel2022} both with OB stars and young open clusters (see their Figs.~12-14).
   
    We go on to consider the pP30 vertical distribution mapped in Fig.~\ref{fig:Zmedian}. Despite of the difference in coordinates projection, it is highly in agreement with both panels of Fig.~5 from \cite{Warp_Merce2019}, which use a sample of OB stars and another one of RGB stars. A very recognisable feature is the positive region around $l\approx60$-$75\,^{\circ}$ that is shown in their two samples as well as in our pP30 sample. The largest positive region at $l\gtrsim130\,^{\circ}$ that starts after $d\approx4$~kpc (i.e. covering a wide field-of-view around the AC beyond $R\approx12$~kpc) for our A-type stars also seems to agree with a combination of both OB and RGB samples from \cite{Warp_Merce2019}: it samples a larger area than the former, while it seems to be shifted about $\sim$~$10\,^{\circ}$ in $l$ with respect to the latter in the sense that the one of the RGB sample is centred at lower Galactic longitudes. The negative clump at $80\,^{\circ}\lesssim l\lesssim130\,^{\circ}$ corresponds better with the OB sample of the aforementioned authors than with their RGB sample, although they coincide for those Galactic longitudes nearer to $l=90\,^{\circ}$. Consequently, intermediate A-type stars seem to show some features detected both with younger and older stars, even if the exact distribution is highly dependent on the age of the tracer.

    The kinematic signature of the Galactic warp seen in our Fig.~\ref{fig:XY_vbcorr_EDR3}, which starts being noticeable at $R\gtrsim12$~kpc around the AC and moves at $v_b^{\rm corr}\approx6$-$7\,\rm{km\,s^{-1}}$ for $R\approx14$~kpc, is also found by \cite{Warp_Merce2019} with their two samples. They find that the warp starts at $R\approx10$-$11$~kpc for their RGB sample and at $R\approx12$-$13$~kpc for their OB sample which, together with our intermediate values, confirms an age dependency of its starting radius. It is also in agreement with the $5$-$6\,\rm{km\,s^{-1}}$ motion found at $R\approx14$~kpc by \cite{WarpUMSGaiaDR2_Poggio2018} for a sample of upper main sequence stars (mainly O, B and A) selected with 2MASS \citep{2MASS_Skrutskie2006} and {\it Gaia} DR2 data \citep{GaiaDR2_2018}. As well as with \cite{AssymDiscDR3_Drimmel2022}, which uses $V_Z$ thanks to the new {\it Gaia} DR3 line-of-sigh velocities and finds a vertical motion of $V_Z(R\approx14$~kpc$)\approx7\,\rm{km\,s^{-1}}$ for a RGB sample (their OB sample has an extremely limited extension and does not reach the required distances).
    
    We analysed a region with downward mean $v_b^{\rm corr}$ found at $(X_{\rm Gal},\,Y_{\rm Gal})\approx(-6,\,6)$~kpc in Fig.~\ref{fig:XY_vbcorr_EDR3}. Its extension to larger heliocentric distances is also detected with OB stars in \cite{Warp_Merce2019}, whose proper motions in $b$ are referred to the local standard of rest instead of to the Sun.
    The map of vertical velocity $V_Z$ of our pP30-RV sample (so, including the full 3D velocities) also presents a similar feature at the same direction beyond $d\approx2$~kpc (bottom panel of Fig.~\ref{fig:VrVphiVz-maps_DR3}). This fact proves that this kinetically coherent structure is real, despite it can be enhanced in Fig.~\ref{fig:XY_vbcorr_EDR3} due to the inherent discrepancy between $v_b^{\rm corr}$ and $V_Z$ combined with extinction effects.
    The $V_Z$ maps from \cite{AssymDiscDR3_Drimmel2022} do not show any peculiarity towards these coordinates for RGB stars, while their OB sample -- which is restricted to $d<2$~kpc -- have almost exactly the same behaviour as our A-type stars. This implies that this kinematic perturbation may be associated with young tracers, either because their bluer colours are more affected by extinction, or because it originated in the kinematics of the gas.
    Nevertheless, the upper-left panel of Fig.~23 from the same authors displays two regions with compression breathing modes using their RGB sample. One of them is nearly symmetric around the AC starting at $R\approx11$~kpc (as we detected in Fig.~\ref{subfig:1725l1875_EDR3}), whereas the other one is just on the controversial location of our interest. Thus, even if the detected signatures may be artificially amplified by some observational limitations, we conclude that the Milky Way disc is being partially compressed in some regions.
    
    The global gradient found in the pP30-RV $V_{\phi}$ map in Fig.~\ref{fig:VrVphiVz-maps_DR3} is qualitatively similar to that of the {\it Gaia} DR3 OB sample in \cite{AssymDiscDR3_Drimmel2022} (middle left panel of their Fig.~21), but it has a noticeably different pattern from the $V_{\phi}$ map shown by their sample of RGB stars (obtained from the data of the middle left panel of their Fig.~16 with an adapted colour axis range).
    The sudden decrease of $\sim$~$15\,\rm{km\,s^{-1}}$ within $\sim$~$2$~kpc in the AC direction is also found with the youngest tracers of \cite{AC_Teresa2021} (see their Fig.~10), and it is partially seen in the {\it Gaia} DR3 OB sample -- although our sample reaches $\sim$~$1$~kpc further. It can also be detected using the open clusters from \cite{OCgdr3_Hunt2023}. Interestingly, zero-age O-type stars with masers at this location do not show any peculiar variation in circular rotation velocity \citep{SpiralArms_Reid2019, MasersPerseus_Sakai2019}.
    
    On the other hand, both inward-velocity features of our pP30-RV $V_R$ map (upper panel in Fig.~\ref{fig:VrVphiVz-maps_DR3}) coincide with the pattern seen by OB stars \citep{AssymDiscDR3_Drimmel2022} and masers \citep{SpiralArms_Reid2019}.
    However, our sample also shows a region with outward $V_R$ at $R>9$~kpc and $5\,^{\circ}\lesssim\phi\lesssim15\,^{\circ}$ -- namely, around $(X_{\rm Gal},Y_{\rm Gal})=(-9.5,1.5)$~kpc -- that is not shown by any of the tracers mentioned in here or above.
    
    Globally, both the OB and the A-type stars show concordant kinematic perturbations that deviate from axisymmetry, mainly in the vertical and the Galactocentric radial velocities. They may be originated by spiral arms, interactions with satellites or bar resonances among others.
    The Galactic latitude limit of the pP30-RV sample highly reduces the statistics very close to the Sun. By contrast, our sample of A-type stars reaches slightly deeper than the one of OB stars. Complementing current samples with the southern Galactic disc from VPHAS+ data \citep{VPHAS_Drew2013} and with WEAVE line-of-sight velocities will highly improve these velocity maps. In turn, this will allow us to understand the dynamics of the Milky Way disc through the connection between these two kind of tracers.
    
   The velocity space substructure found towards the AC using velocities in Galactic coordinates directions (Fig.~\ref{fig:VZvsVphi_pP50}) is highly in agreement with Fig.~13 from \cite{AC_Teresa2021} and with the results at the AC shown in Fig.~7 from \cite{Disturbances_McMillan2022}. However, even if detected inhomogeneities correspond in most of the cases, our sample (without mixing different stellar populations) shows them as better-defined clumpy regions and with a higher degree of detail. In consequence, our A-type stars selection reveals some substructure that had previously gone unseen, such as the presence of a bimodality for $Z_{\rm Gal}>0$ and a trimodality for $Z_{\rm Gal}<0$ between $R=11$~kpc and $R=15$~kpc.

\section{Conclusions}\label{sectionConclusions}
    We constructed an extended sample of A-type stars to unveil Milky Way disc density structures and kinematic perturbations. It was selected using a colour-colour diagram from IGAPS photometric bands and includes {\it Gaia} DR3 astrometry and line-of-sight velocities. Our main findings are as follows.
    
    \begin{itemize}
    
        \item We present a catalogue of 3\,532\,751 A-type stars up to magnitude $r=19$~mag. When compared with a spectroscopic sample from LAMOST, the contamination -- mostly from slightly colder populations -- is estimated to be 10\%. When dereddening the subsample with relative error in parallax smaller than 30\%, our IGAPS selection criterion retains less than 1\% of white dwarfs or giants in the sample, and the fraction of stars with derived absolute magnitudes fainter than that of F0-type stars is about 25\% (two-thirds of them being compatible with F-type stars).
        The catalogue has 31\,934 sources with line-of-sight velocities from {\it Gaia} DR3.
        
        \item We detect stellar density enhancements associated with the Perseus arm at Galactocentric coordinates $(X_{\rm Gal},Y_{\rm Gal})\approx(-10.00, \, 1.75)$~kpc, with the Cygnus region at $(X_{\rm Gal},Y_{\rm Gal})\approx(-8.0, \, 1.0)$-$(-7.5, \, 2.0)$~kpc and with the Cepheus Spur at $(X_{\rm Gal},Y_{\rm Gal})\approx(-9.00, \, 0.75)$~kpc. We also find a low-density region at $(X_{\rm Gal},Y_{\rm Gal})\approx(-10.0, \, 0.5)$~kpc already observed using open clusters, upper main sequence stars, and masers.
        
        \item The analysis of the vertical distribution of stellar densities proves that many of the prominent differences between $b>0\,^{\circ}$ and $b<0\,^{\circ}$ are caused by extinction and might not entirely correspond with real asymmetric distributions.
        The imprint of the Galactic warp is not clear considering just the density distribution of the sample.
        
        \item The cylindrical component of the velocity $V_{\phi}$ presents large variations with values above $240\,\rm{km\,s^{-1}}$ in the first quadrant and a rapidly decreasing trend in the anticentre (AC) direction with values ranging from $240\,\rm{km\,s^{-1}}$ at the Sun position to $225\,\rm{km\,s^{-1}}$ at $R\approx10$~kpc.
        
        \item The Galactocentric radial velocity $V_R$ shows patterns that deviate up to 10-20$\,\rm{km\,s^{-1}}$ from circular orbits. The main features are two regions with mean $V_R$ towards the Galactic centre (GC) that are also seen with other tracers and an outward-velocity region around $(X_{\rm Gal},Y_{\rm Gal})=(-9.5,1.5)$~kpc that is not detected in other studies.
        
        \item We established a simple model (Appendix~\ref{appendixKinModel}) and determined that the velocity in the Galactic latitude direction corrected by the solar motion ($v_b^{\rm corr}$) is practically equivalent to the Galactocentric vertical velocity ($V_Z$) along the GC-AC direction as long as $V_R\approx0\,\rm{km\,s^{-1}}$. On the other hand, $v_b^{\rm corr}$ and $V_Z$ are extremely discrepant towards other Galactic longitudes, where their difference has a strong dependence on $Z_{\rm Gal}$ that creates artificial breathing modes.
        
        \item Using both $v_b^{\rm corr}$ on the sample with proper motions and also $V_Z$ on the reduced 6D sample, we detected a kinetically peculiar structure towards the $60\,^{\circ}\leq l\leq 75\,^{\circ}$ direction that moves downwards. It reaches $V_Z$ about $-4\,\rm{km\,s^{-1}}$ for our A-type stars. This behaviour is also shown by OB stars, but is more elusive for the older RGB stars. Confirming it as a young kinematic structure.
        
        \item We find that the kinematic signature of the Galactic warp begins at $R\!\approx\!12$~kpc ($d\approx4$~kpc) and that it has an amplitude of $v_b^{\rm corr}\approx6$-$7 \,\rm{km\,s^{-1}}$ at $R\approx14$~kpc. Our result favour the scenario where the starting radius of the warp changes with the age of the tracer in the sense that it begins further for younger samples.
        
        \item We also show that A-type stars have a very inhomogeneous and asymmetric $V_Z\equiv v_b^{\rm corr}$ vs $V_{\phi}\equiv -v_l^{\rm corr}$ velocity space, proving that these peculiarities are shared among different tracers. In particular, our sample allow us to trace two (three) major overdensities at $Z_{\rm Gal}>0$ ($Z_{\rm Gal}<0$) from $R=11$~kpc up to $R=15$~kpc not shown in previous studies.
        
    \end{itemize}

    Thanks to {\it Gaia} and large ongoing photometric surveys, this work upgrades A-type stars as a new population for large-scale Galactic disc structure and kinematics studies. We demonstrate that A-type stars are very powerful tracers that can be used in addition to those that have historically been used as an age transition to complete and expand our knowledge of the properties and the evolution of the Milky Way.
    

\begin{acknowledgements}
    We would like to thank the anonymous referee for the comments and suggestions, which greatly improved this work. We would also like to thank Dr.~Luis Aguilar and Dr.~Teresa Antoja for our scientific discussions and comments, as well as David Altamirano for the exchange of ideas and some programming codes we have had.\\
    
    This work made use of TOPCAT\footnote{\url{http://www.star.bris.ac.uk/~mbt/topcat/}} and STILTS\footnote{\url{http://www.star.bris.ac.uk/~mbt/stilts/}} Visual Observatory tools.\\
    
    This research made use of Astropy,\footnote{\url{http://www.astropy.org}} a community-developed core Python package for Astronomy \citep{astropy:2013, astropy:2018}.\\
    
    This work has made use of data from the European Space Agency (ESA) mission {\it Gaia} (\url{https://www.cosmos.esa.int/gaia}), processed by the {\it Gaia} Data Processing and Analysis Consortium (DPAC, \url{https://www.cosmos.esa.int/web/gaia/dpac/consortium}). Funding for the DPAC has been provided by national institutions, in particular the institutions participating in the {\it Gaia} Multilateral Agreement.\\

    This work was part of the PRE2021-100596 grant funded by Spanish MCIN/AEI/10.13039/501100011033 and by ESF+. It was also (partially) funded by "ERDF A way of making Europe" by the “European Union” through grants RTI2018-095076-B-C21 and PID2021-122842OB-C21, and the Institute of Cosmos Sciences University of Barcelona (ICCUB, Unidad de Excelencia ’Mar\'{\i}a de Maeztu’) through grant CEX2019-000918-M.
\end{acknowledgements}


\bibliography{references}


\begin{appendix}

\onecolumn
\section{Catalogue columns}\label{appendixColumns}
The catalogue of 3\,532\,751 A-type stars presented in this work is available through CDS. It contains photometry from IGAPS, as well as astrometry and radial velocities from {\it Gaia} DR3. It also includes distances estimated with the EDSD prior explained in Sect.~\ref{sectionData}, coordinates and velocities computed as described in Sect.~\ref{sectionDefinitions}, and {\it Gaia} G-band and \bprp colour extinctions (the two last ones, for those stars of the pP30 subsample that have {\it Gaia} \bprp colour).
\begin{longtable}{r c c l}
\caption{List of columns of the catalogue, their units, and descriptions.} \label{tab:columns} \\
\hline
  & Column & Units & Description \\
\hline
1  &           name          &         ---         &  IGAPS source designation. \\
2  &           r\_I          &         mag         &  IGAPS r-band magnitude. \\
3  &         rErr\_I         &         mag         &  IGAPS r-band magnitude uncertainty. \\
4  &            ha           &         mag         &  IGAPS $H\alpha$-band magnitude. \\
5  &          haErr          &         mag         &  IGAPS $H\alpha$-band magnitude uncertainty. \\
6  &            i            &         mag         &  IGAPS i-band magnitude. \\
7  &           iErr          &         mag         &  IGAPS i-band magnitude uncertainty. \\
8  &        source\_id       &         ---         &  {\it Gaia} DR3 source identifier. \\
9  &    phot\_g\_mean\_mag   &         mag         &  {\it Gaia} DR3 G-band apparent magnitude. \\
10 &          bp\_rp         &         mag         &  {\it Gaia} DR3 \bprp colour. \\
11 &            ra           &         deg         &  Right ascension from {\it Gaia} DR3. \\
12 &        ra\_error        &         mas         &  Right ascension error from {\it Gaia} DR3. \\
13 &           dec           &         deg         &  Declination from {\it Gaia} DR3. \\
14 &        dec\_error       &         mas         &  Declination error from {\it Gaia} DR3. \\
15 &         parallax        &         mas         &  {\it Gaia} DR3 parallax. \\
16 &     parallax\_error     &         mas         &  Error of {\it Gaia} DR3 parallax. \\
17 &           pmra          & $\rm{mas\,yr^{-1}}$ &  Proper motion in right ascension direction from {\it Gaia} DR3. \\
18 &       pmra\_error       & $\rm{mas\,yr^{-1}}$ &  Error of proper motion in right ascension direction from {\it Gaia} DR3. \\
19 &          pmdec          & $\rm{mas\,yr^{-1}}$ &  Proper motion in declination direction from {\it Gaia} DR3. \\
20 &      pmdec\_error       & $\rm{mas\,yr^{-1}}$ &  Error of proper motion in declination direction from {\it Gaia} DR3. \\
21 &          ruwe           &         ---         &  {\it Gaia} DR3 renormalised unit weight error. \\
22 &     radial\_velocity    &  $\rm{km\,s^{-1}}$  &  {\it Gaia} DR3 radial velocity. \\
23 & radial\_velocity\_error &  $\rm{km\,s^{-1}}$  &  {\it Gaia} DR3 radial velocity error. \\
24 &           dist          &         kpc         &  Distance to the Sun computed according to an EDSD prior with a length scale of 3kpc. \\
25 &           Xgal          &         kpc         &  Galactocentric X coordinate. \\
26 &           Ygal          &         kpc         &  Galactocentric Y coordinate. \\
27 &           Zgal          &         kpc         &  Galactocentric Z coordinate. \\
28 &            R            &         kpc         &  Cylindrical Galactocentric radial coordinate. \\
29 &           phi           &         deg         &  Cylindrical Galactocentric azimuthal coordinate. \\
30 &           pml           & $\rm{mas\,yr^{-1}}$ &  Proper motion in Galactic longitude direction. \\
31 &           pmb           & $\rm{mas\,yr^{-1}}$ &  Proper motion in Galactic latitude direction. \\
32 &         vb\_corr        &  $\rm{km\,s^{-1}}$  &  Velocity in Galactic longitude direction corrected for solar motion. \\
33 &           V\_R          &  $\rm{km\,s^{-1}}$  &  Cylindrical Galactocentric radial velocity. \\
34 &          V\_phi         &  $\rm{km\,s^{-1}}$  &  Cylindrical Galactocentric azimuthal velocity. \\
35 &           V\_Z          &  $\rm{km\,s^{-1}}$  &  Cylindrical Galactocentric vertical velocity. \\
36 &           A\_G          &         mag         &  G-band extinction derived from \cite{Bayestar_Green2019} dustmap. \\
37 &       E\_BPminusRP      &         mag         &  \bprp reddening derived from \cite{Bayestar_Green2019} dustmap. \\
\end{longtable}
\twocolumn

\section{Extinction law in {\gdr3}}\label{appendixExtinction}
    \cite{Ramos2020} provided some extinction law transformations to estimate the absorption in {\gaia} $G$ passband, $A_G$, when we know the absorption in the $V$ Johnson band, $A_V$, and the \bprp colour of the star in {\gdr2}. We update here this relationship using the new passband transmissivity derived in {\gedr} \citep{Riello2021}. These transmissivity curves are also valid for photometric data in {\gdr3}. Besides, we also provide a new relationship useful to predict the colour excess in \bprp, $E(\bprp)$, once $A_G$ and \bprp are known. 
    
    The fitted laws were obtained using simulations derived from BaSeL-3.1 \citep{basel31} spectral energy distributions, using the same methodology as in \cite{Jordi2010}. For the Johnson $V$ passbands, we used the response curve provided by \cite{bessell2012}.
    
    The extinction law by \cite{Cardelli1989} was used to simulate reddened sources with $0<A_{550}<11$~mag, where $A_{550}$ is the monochromatic extinction magnitude at wavelength $\lambda=550$~nm.
    
\begin{figure}
        \centering
        \begin{subfloat}
        \centering
            \includegraphics[width=0.97\linewidth]{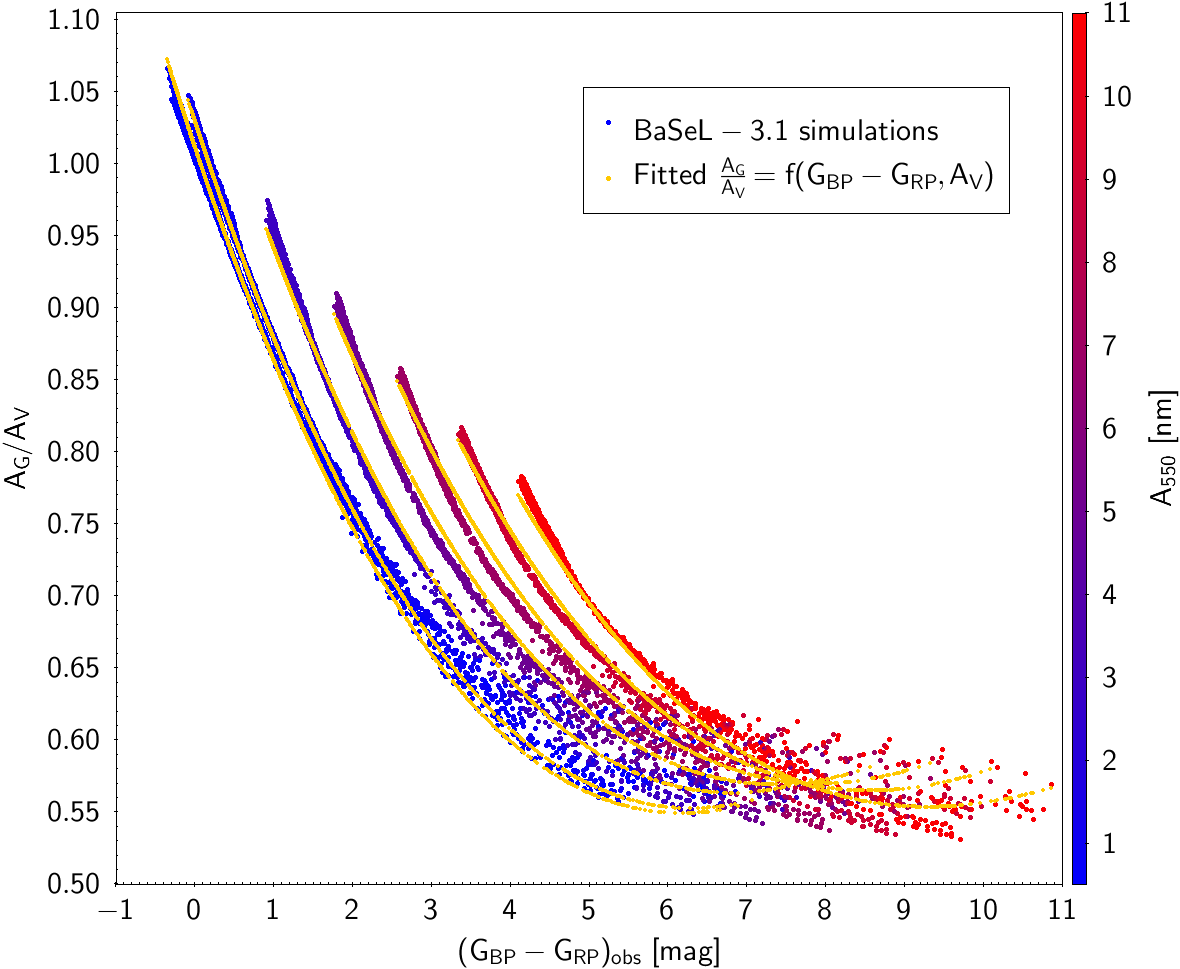}
        \end{subfloat}
        \vfill
        \begin{subfloat}
        \centering
            \includegraphics[width=0.97\linewidth]{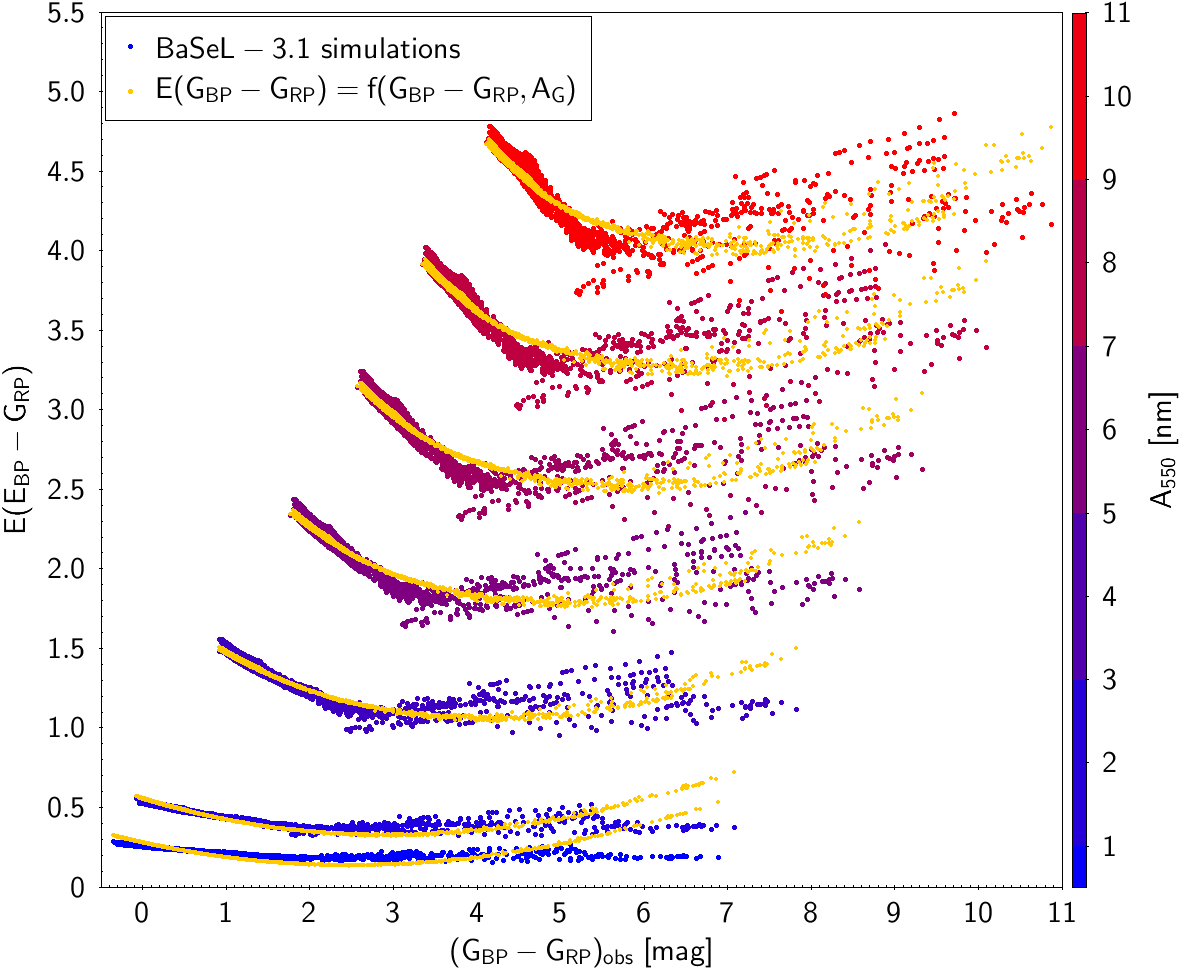}
        \end{subfloat}
        \caption{Fitted relationship between ${A_G}/{A_V}$ (left) or $E(\bprp)$ (right) as a function of the observed $G_{\rm BP}-G_{\rm RP}$ colour (affected by reddening) derived from simulations using BaSeL-3.1 synthetic spectral energy distributions \citep{basel31} and {\gedr} passbands from \citet{Riello2021}.}
        \label{fig:AGAVvsBPRP}
    \end{figure}

    The fitted laws consider a third degree polynomial with the observed \bprp (affected by reddening) and linear relationship with the absorption ($A_V$ for $A_G/A_V$ relationship and $A_G$ for $E(\bprp)$ relationship). We also consider a crossed term multiplying the absorption and the colour of the source. The obtained fitted relationships are included in Eqs.~\ref{eq:extinction_transformation1} and \ref{eq:extinction_transformation2} and are plotted in Fig.~\ref{fig:AGAVvsBPRP}.

    \begin{eqnarray}
        \label{eq:extinction_transformation1}
        \nonumber A_G/A_V &=& 0.99431 -0.16269 (\bprp) \\
        \nonumber &&+0.017012 (\bprp)^2 \\
        &&-0.00039535 (\bprp)^3 \\
        \nonumber &&+0.035811 A_V \\
        \nonumber &&-0.0046467  (\bprp) A_V \\
        \label{eq:extinction_transformation2}
        \nonumber E(\bprp) &=& 0.013302 -0.078092 (\bprp) \\
        \nonumber &&+0.021183 (\bprp)^2 \\
        &&-0.00034409 (\bprp)^3 \\
        \nonumber &&+0.52634  A_G \\
        \nonumber &&+0.0037629 (\bprp) A_G 
    \end{eqnarray}
    
    Figure~\ref{fig:AGAVvsBPRP} shows that the validity of the law to obtain $E(\bprp)$ is limited for very red and reddened sources, but it can provide a reasonable estimations for other kind of sources.

\section{Estimation of the completeness distance}\label{appendixCompleteness}
    In this section we describe a method to estimate the sample's completeness by computing the furthest distance $d_{\rm lim}$ at which an A5 star can be observed given a limiting magnitude $r_{\rm lim}=19$~mag. We followed these steps:
    \begin{enumerate}
        \item We created a grid of 30~arcmin in $b$ and 10~arcmin in $l$.
        \item For each direction, we obtained the $E(B-V)$ reddening from \cite{Bayestar_Green2019} dustmap at distance steps of 100~pc.
        \item We transformed these $E(B-V)$ values to PANSTARRS r-band \citep{PANSTARRSfiltersLim_Hodapp2004, PANSTARRSfiltersTrans_Magnier2007} extinction\footnote{Using \cite{A4000toAr_Sale2015} transformations and \cite{ExtinctionMap_Sale2014} extinction map, we checked that differences between $A_r$ in IPHAS r-band and $A_r$ in PANSTARRS r-band do not exceed 0.04~mag in our sky coverage, which leads to distance uncertainties smaller or of the order of 1~pc. Thus, both extinctions are equivalent for our purpose.} according to $A_{r \ \rm{IPHAS}}\approx A_{r \ \rm{PANSTARRS}}= 2.271~E(B-V)$ \citep{ExtinctionLaw_Fitzpatrick2007}. Results obtained using extinction values from \cite{ExtinctionMap_Sale2014} only have minor differences with respect to those derived from \cite{Bayestar_Green2019}.
        \item In order to obtain a distribution of $A_r$ as a function of distance and Galactic longitude, we integrated the aforementioned grid of extinctions within bins of $2\,^{\circ}$ in $l$ by selecting the 84th percentile (equivalent to the median plus 1$\sigma$) of all included directions.
        \item We then computed the corresponding apparent magnitude $r$ that this tracer (an A5 star, with assumed absolute r-band magnitude of $M_r=1.844$~mag) would have at each Galactic longitude bin and distance.
        \item Finally, we selected the furthest distance bin, $d_{\rm lim}$, for which $r$ is smaller than $r_{\rm lim}=19$~mag.
    \end{enumerate}
    
    This definition of $d_{\rm lim}$ is just an estimation, and the sample can lack stars at nearer distances mainly because of crowding or saturation. Moreover, when using a sample with additional cuts (e.g. some $\sigma_{\varpi} / \varpi$ restriction), this approach useful for magnitude-limited samples becomes oversimplified.
    In this second case, the determination of the completeness limits becomes much more complex and may require the usage of models or simulations. Nonetheless, this simple procedure gives $d_{\rm lim}$ values that generally agree with distances at which dark features become yellowish in the left panel of Fig.~\ref{fig:structureXYmaps}; except for around the AC, where they become much larger than 6~kpc.
    
    The values of $d_{\rm lim}$ for the pP30 sample split into two vertical bins are shown as a dotted orange ($b<0\,^{\circ}$) and a dashed blue ($b>0\,^{\circ}$) lines in Figs.~\ref{fig:SupDens_bNPdiff} and \ref{fig:Zmedian}.

\section{Kinematic model and biases}\label{appendixKinModel}
    When $v_{\rm los}$ is missing, the transformation from heliocentric velocities $(v_l^{\rm uncorr},\,v_b^{\rm uncorr},\,v_{\rm los})$ into Galactocentric $(V_R,\,V_{\phi},\,V_Z)$ cannot be computed. A frequently used approximation is to consider that $v_b^{\rm corr}$ is similar to $V_Z$ for stars with small enough Galactic latitudes, as $b$ and $Z_{\rm Gal}$ directions are almost parallel in this case. However, the projection of the true $(V_R,\,V_{\phi},\,V_Z)$ in the $b$ direction has a term that depends on $V_R$ and $V_{\phi}$ apart from the one proportional to $V_Z{\rm cos}(b)$. As proven by \cite{LuisVbBias_Croswell1987} in their Fig.~4, this additional component makes $v_b^{\rm corr}$ a low-accuracy estimator of $V_Z$ whose fidelity changes with $V_R$ and $V_{\phi}$, as well as with the three-dimensional position of the target within the Galaxy. Here we use a model to describe how $v_b^{\rm corr}$ behaves in comparison with $V_Z$.
    
    The model is based on a grid in $(l,\,b,\,d)$ that covers all Galactic longitudes in steps of $1\,^{\circ}$, has the same Galactic latitude range as the sample ($|b|\leq5\,^{\circ}$) with a sampling interval of $0.1\,^{\circ}$, and reaches up to $d=10$~kpc with a grid-spacing of 0.2~kpc. Velocities are forced to be in circular orbits ($V_{R,\:\rm{Model}}=0\,\rm{km\,s^{-1}}$) and to a flat rotation curve with $V_{\phi,\:\rm{Model}}=240\,\rm{km\,s^{-1}}$ \citep{Vphi240_Huang2016}. In turn, the vertical motion of each point of the grid is modelled as a vertical oscillation with velocity amplitude $|V_{Z,\:\rm{Model}}(Z_{\rm Gal})|=V_{Z,0}\,|\rm{sech}^2(Z_{\rm Gal}/Z_0)|$. The scale height $Z_0$ is chosen to be 200~pc as for the stellar surface density (Sect.~\ref{sectionXYdistrib}), and the maximum modulus $V_{Z,0}$ equals $10\,\rm{km\,s^{-1}}$ to be comparable with the dispersion in vertical velocities found by \cite{AC_Teresa2021}. The orientation of this motion is assigned randomly for each grid point. The proper motion in Galactic latitude is computed from this modelled velocity field and it is then corrected for solar motion according to Eq.~\ref{eq:v_corr}. This representation does not intend to reproduce in detail the real kinematics of the Milky Way. Instead, it is applied to enable the study of vertical biases.
    
\begin{figure}
        \centering
        \begin{subfloat}
        \centering
            \includegraphics[width=0.9\hsize]{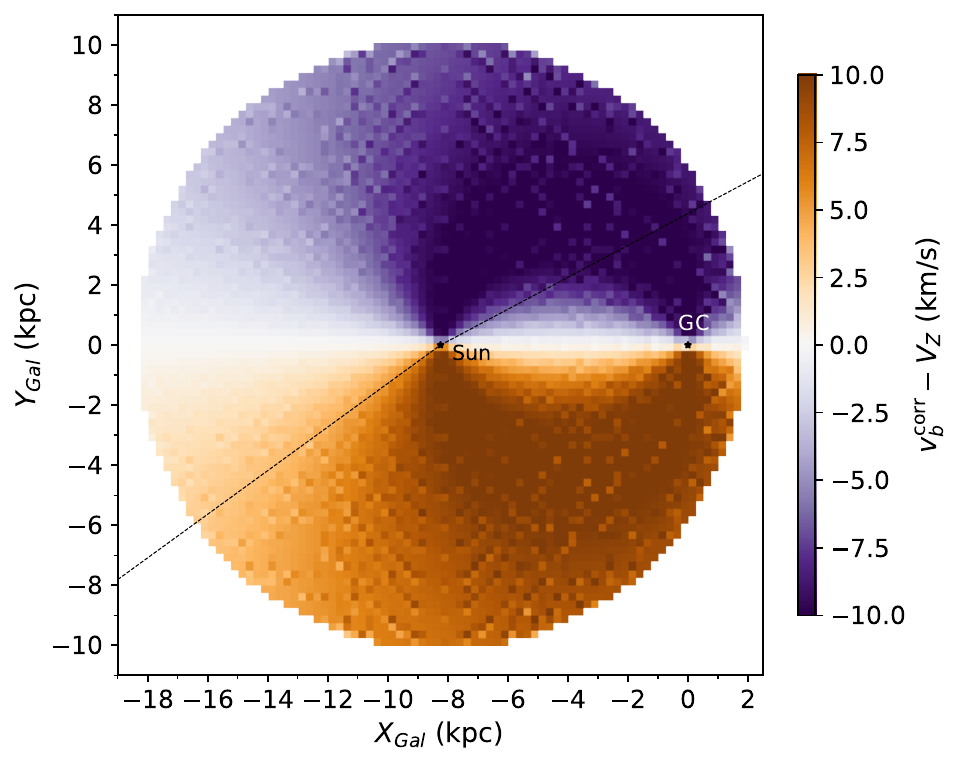}
        \end{subfloat}
        \vfill
        \begin{subfloat}
        \centering
            \includegraphics[width=0.9\hsize]{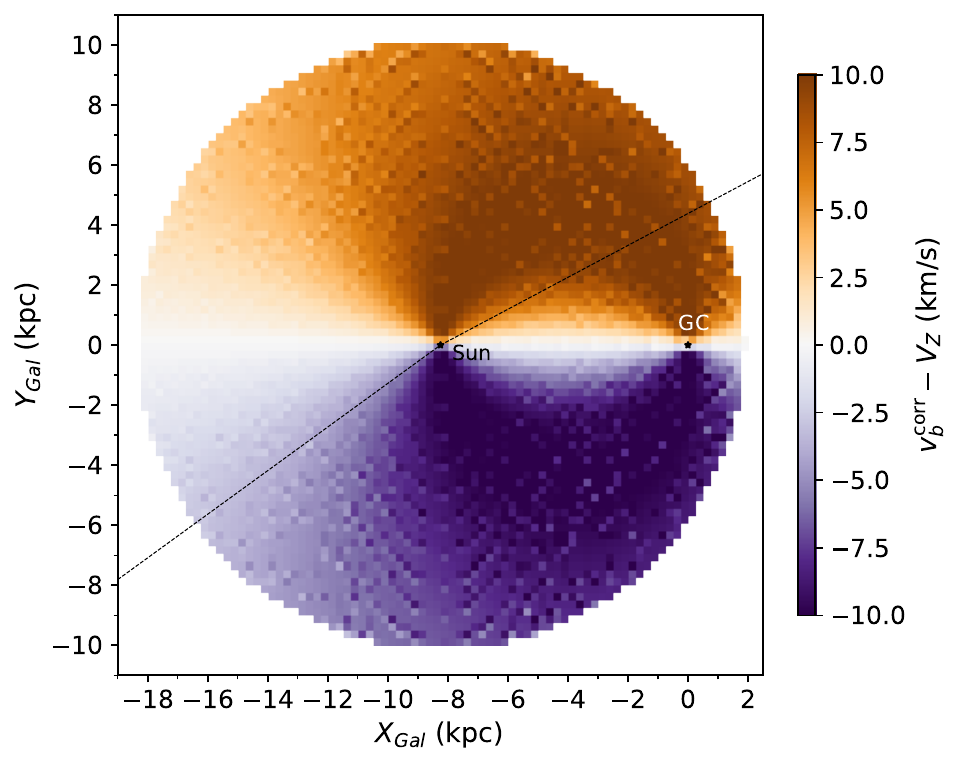}
        \end{subfloat}
        \caption{XY map of median differences between $v_b^{\rm corr}$ and $V_Z$ obtained with the model restricted to $Z>0$ (up) and to $Z<0$ (bottom). In both panels, the GC is at the origin of coordinates and the Sun is centred. Bins are squares of side 0.25~kpc. Black diagonal lines show the Galactic longitude limits of our samples.}
        \label{fig:XYnorthsouth_abs_vbcorr-W_Model2}
    \end{figure}

    To quantify the discrepancy among $v_b^{\rm corr}$ and $V_Z$, we plot the distribution of the median $v_b^{\rm corr}-V_Z$ for $Z>0$ and for $Z<0$ separately in Fig.~\ref{fig:XYnorthsouth_abs_vbcorr-W_Model2}.
    For a set of stars having perfectly circular orbits around the GC -- like those in the model --, line-of-sight velocities at exactly $l=0\,^{\circ}$ and $l=180\,^{\circ}$ would be null at $b=0\,^{\circ}$. Thus, towards the GC and the AC, the three-dimensional velocity space would be fully recovered with no more than proper motions. In addition, both the cylindrical Galactocentric and the Galactic heliocentric coordinates systems have parallel axes along these directions. These are the reasons for which no deviations are found there.
    This figure proves that $v_b^{\rm corr}-V_Z$ grows in absolute value as $l$ deviates from the AC following a non-trivial pattern that has two singularities because of the combination of a spherical and a cylindrical reference systems with different origins. Additionally, this pattern has opposite signs for $Z>0$ and $Z<0$, which creates -- for any heliocentric distance -- an artificial compression (expansive) breathing mode in $v_b^{\rm corr}$ for $0\,^{\circ}\lesssim l\lesssim 180\,^{\circ}$ ($180\,^{\circ}\lesssim l\lesssim 360\,^{\circ}$) that does not exist in $V_Z$.

    For the imposed $V_{\phi,\:\rm{Model}}=240\,\rm{km\,s^{-1}}$, vertically-averaged differences between $V_Z$ and $v_b^{\rm corr}$ reach up to $\sim$~$10\,\rm{km\,s^{-1}}$. Consequently, $v_b^{\rm corr}$ and $V_Z$ can become extremely discrepant outside the AC, the GC or the plane $b=0\,^{\circ}$.
    While this effect could in principle be corrected for each star knowing its $(l,\,b,\,d)$ coordinates and according to some $(V_R,\,V_{\phi},\,V_Z)$ modelling, the ignorance of the real Galactic velocities can introduce undesired systematic errors in the $v_b^{\rm corr}-V_Z$ correction and so, we decided not to apply it. We compared several values for $V_{\phi,\:\rm{Model}}$ and $V_{Z,\:\rm{Model}}(Z_{\rm Gal})$ finding that maximum $|v_b^{\rm corr}-V_Z|$ are always obtained for the most extreme $b$ around the ring linking the Sun and the GC (as in Fig.~\ref{fig:XYnorthsouth_abs_vbcorr-W_Model2}) and that they are of the order of 18-20$\,\rm{km\,s^{-1}}$. When varying $V_{\phi,\:\rm{Model}}$ between 220 and 260$\,\rm{km\,s^{-1}}$, differences in the $v_b^{\rm corr}-V_Z$ correction reach 3-4$\,\rm{km\,s^{-1}}$. Combining this with different approaches for $V_{Z,\:\rm{Model}}$ only produces minor variations at the level of $<0.1\,\rm{km\,s^{-1}}$. When including a non-vanishing radial velocity $V_{R,\:\rm{Model}}=\pm15\,\rm{km\,s^{-1}}$ in the model, the shape of the Fig.~\ref{fig:XYnorthsouth_abs_vbcorr-W_Model2} pattern is no longer symmetric with respect to the GC-AC line; where the $v_b^{\rm corr}-V_Z$ corrections reach the larger differences ($\lesssim3\,\rm{km\,s^{-1}}$) among $V_{R,\:\rm{Model}}$ modellings.
    The exact range of $|v_b^{\rm corr}-V_Z|$ and the shape of its XY pattern slightly depends on $(U_{\odot}, \, V_{\odot}, \, W_{\odot})$, even if there are no appreciable changes when varying them within a reasonable range.
    
    In addition to its Galactic longitude dependency, $v_b^{\rm corr}-V_Z$ is also highly dependent on $Z_{\rm Gal}$. In other words, this deviation changes when getting away from the Galactic plane ($Z_{\rm Gal}=0$). Since this dependence is symmetric with respect to it, the effect is compensated when averaging over all $Z_{\rm Gal}$ inside bins in the XY plane. However, vertical asymmetries in stellar density (e.g. due to extinction) would make this effect relevant because of the non-uniform sampling.
    
\begin{figure}
        \centering
        \begin{subfigure}[b]{0.48\hsize}
        \centering
            \includegraphics[width=\hsize]{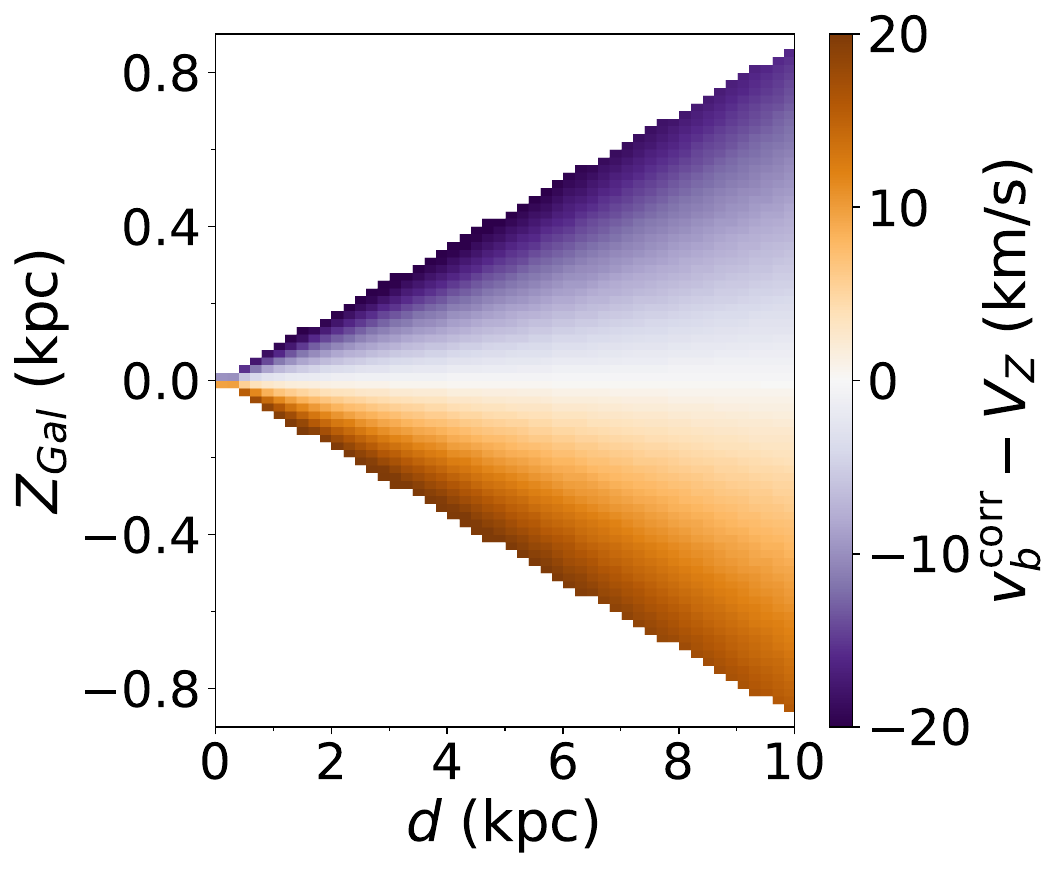}
            \caption{$60 \,^{\circ} \leq l \leq 75 \,^{\circ}$}
            \label{subfig:60l75_Model2}
        \end{subfigure}
        \hfill
        \begin{subfigure}[b]{0.51\hsize}
        \centering
            \includegraphics[width=\hsize]{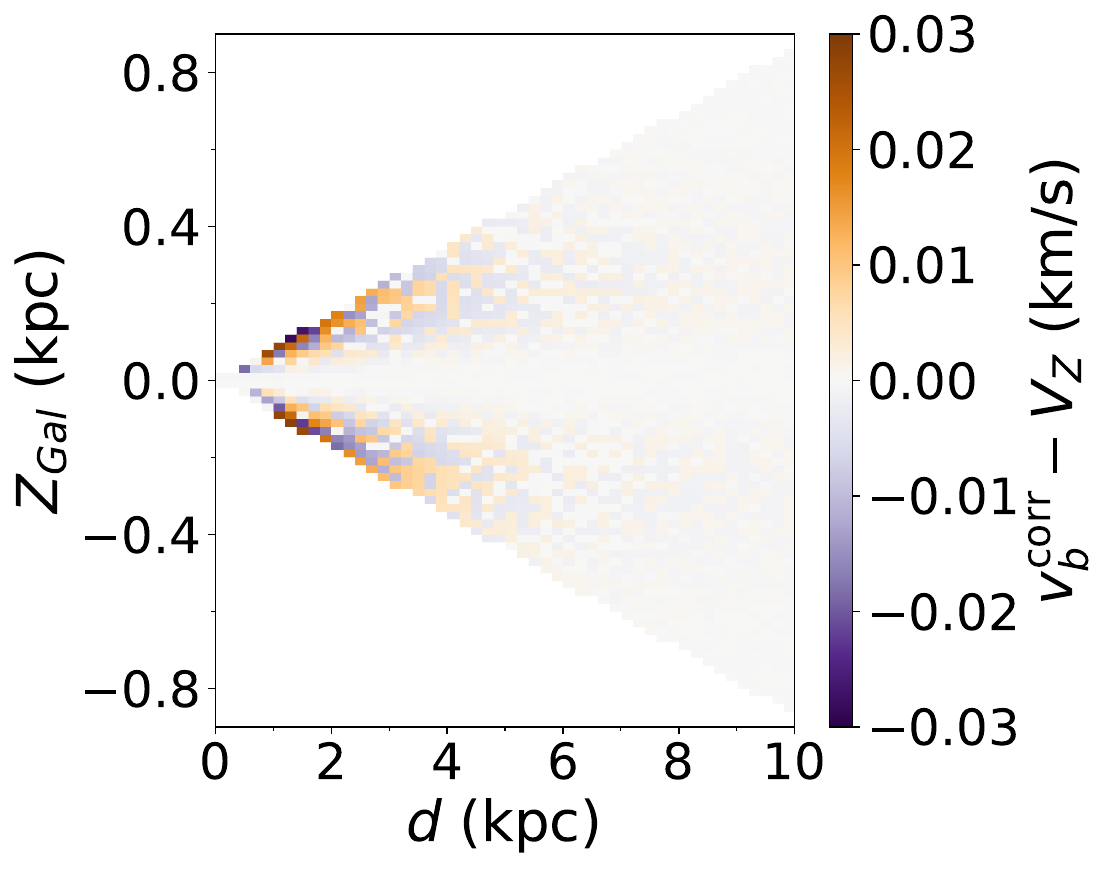}
            \caption{$172.5 \,^{\circ} \leq l \leq 187.5 \,^{\circ}$}
            \label{subfig:1725l1875_Model2}
        \end{subfigure}
        \caption{Vertical Galactic coordinate as a function of heliocentric distance, colour-coded by the median of the modelled difference $v_b^{\rm corr}-V_Z$. Panel (a) shows the slice between $60\,^{\circ}\leq l\leq75\,^{\circ}$ while panel (b) contains grid points with $172.5\,^{\circ}\leq l\leq187.5\,^{\circ}$. In both cases, the size of the bins is 0.2~kpc in $d$ and 20~pc in $Z_{\rm Gal}$. Notice that both colour scales have very different ranges.}
        \label{fig:Zd_vbcorrVZ_Model2}
    \end{figure}

    Figure~\ref{fig:Zd_vbcorrVZ_Model2} demonstrates this effect for two particular slices in $l$ that are of special interest for the observational sample (see Sect.~\ref{subsectionKinematicsNoRV} and compare with Fig.~\ref{fig:Zd_vbcorrVZ_EDR3}). Both cuts are chosen so that they have the same width in $l$, being defined by $60\,^{\circ}\leq l\leq 75\,^{\circ}$ and $172.5\,^{\circ}\leq l\leq 187.5\,^{\circ}$. The former encompasses a kinetically peculiar structure found in Fig.~\ref{fig:XY_vbcorr_EDR3} (Sect.~\ref{subsectionKinematicsNoRV}); whereas the latter is symmetric around the AC and proves that this is a privileged direction where $v_b^{\rm corr}\equiv{V_Z}$ is verified assuming almost circular orbits.
    
    In conclusion, using $v_b^{\rm corr}$ as the substitute of $V_Z$ should be treated with care, always verifying that trends found cannot be reproduced by biases.
    
\end{appendix}

\end{document}